\documentclass[aps,prd,onecolumn,superscriptaddress,showpacs,nofootinbib,floatfix]{revtex4}

\usepackage{dcolumn}
\usepackage{bm}
\usepackage[dvips]{graphicx}
\usepackage{float}
\usepackage{epsfig}
\usepackage{graphicx}
\usepackage{longtable} 
\usepackage{rotating}
\usepackage{amsmath}
\usepackage{amsfonts}
\usepackage{amssymb}

\newcommand{\sqrts}{\sqrt{s}}
\newcommand{\pp}{$p$-$p$}
\newcommand{\ppbar}{$p$-$\bar p$}
\newcommand{\qg}{$q g$}
\newcommand{\qqbar}{$q \bar q$}

\newcommand{\GeVcc}{GeV/c$^2$}

\newcommand{\ETg}{E_{_T}^{^\gamma}}

\newcommand{\jetphox}{{\sc jetphox}}
\newcommand{\pythia}{{\sc pythia}}

\newcommand{\lhapdf}{{\sc lhapdf}}
\newcommand{\dzero}{D$\emptyset$}

\def\mean#1{\ensuremath{\left<#1\right>}}

\def\cO#1{{{\cal{O}}}\left(#1\right)}

\begin{document}
%\hfill {\sf ICCUB-10-0ii}
%\vspace{0.5cm}

%\title{Sensitivity of isolated photon production at high transverse momentum at 
%\title{Sensitivity of high transverse momentum isolated photon production at 
\title{Sensitivity of isolated photon production at TeV hadron \\
%the Tevatron and\\ Large Hadron Collider 
colliders to the gluon distribution in the proton}

\author{Rapha\"elle Ichou}
\affiliation{SUBATECH, 4 rue Alfred Kastler, BP 20722, 44307 Nantes Cedex 3, France}
\author{David~d'Enterria}
\affiliation{ICREA \& ICC-UB, Universitat de Barcelona, 08028 Barcelona, Catalonia}

\begin{abstract}
\noindent
We compare the single inclusive spectra of isolated photons measured at high transverse energy %($E_{T}$) 
in proton-antiproton collisions at $\sqrts$~=~1.96~TeV with next-to-leading order %(NLO) 
perturbative QCD predictions %obtained with the \jetphox\ code 
with various parametrizations of the parton distribution functions (PDFs). 
%and other theoretical ingredients. A $\chi^2$-analysis indicates that, within 
Within the experimental %(energy scale) and theoretical (scales dependence) uncertainties, 
and theoretical uncertainties, the Tevatron data can be reproduced equally well by the recent 
CTEQ6.6, MSTW08 and NNPDF1.2 PDF sets. %parametrizations of the proton PDFs. 
We present also the predictions for isolated $\gamma$ spectra in proton-proton collisions 
at $\sqrts$~=~14~TeV %at the CERN Large Hadron Collider (LHC) 
for central ($y~=~0$) and forward ($y~=~4$) rapidities relevant for LHC experiments. 
Different proton PDFs result in maximum variations of order $\pm$30\% in the expected 
$\ETg$-differential isolated $\gamma$ cross sections. The inclusion of the isolated photon 
data in global PDF fits will place extra independent constraints on the gluon density.
\end{abstract}


\maketitle

%%%%%%%%%%%%%%%%%%%%%%%%%%%%%%%%%%%%%%%%%%%%
\section{Introduction}

%The study of photon
The production of photons with large transverse energy ($E_{_T}\gg\Lambda_{\rm QCD}\approx$~0.2~GeV)
%in hard parton-parton scatterings is a valuable testing ground of the perturbative regime of Quantum Chromodynamics 
in high-energy hadronic collisions is an interesting process in itself as a testing ground of the perturbative regime of 
Quantum Chromodynamics (pQCD)~\cite{Owens:1986mp}, as well as a possible background for important new physics 
searches (such as e.g. often exemplified in the Higgs boson decay into two high-$E_{_T}$ photons~\cite{Binoth:2002wa}). 
%\cite{Djouadi:2005gi}). 
From a pQCD perspective, prompt photons issuing from hard parton-parton scatterings constitute a particularly 
clean testbed of perturbation theory in the %fixed-order 
collinear-~\cite{ABDFS,GV} and $k_T$-~\cite{kT} factorization (or colour-dipole~\cite{Machado:2008nz,Kopeliovich:2009yw})
approaches as well as of various logarithmic resummation techniques~\cite{nll}. In addition, high-$E_{_T}$ photons also 
yield valuable information about non-perturbative objects such as the parton distribution functions (PDFs) 
in the proton~\cite{Aurenche:1988vi,Vogelsang:1995bg} and the parton-to-photon fragmentation functions (FFs)~\cite{BFG,PFF}. 
%~\cite{Gluck:1992zx,BFG}. 
At lowest order in the electromagnetic and strong coupling constants $\mathcal{O}(\alpha\,\alpha_s)$, three 
partonic mechanisms produce prompt photons in hadronic collisions:
(i) quark-gluon Compton scattering $qg\rightarrow \gamma q$, (ii) quark-antiquark annihilation $q\,\bar q\to\gamma\,g$, 
and (iii) the collinear fragmentation of a final-state parton into a photon\footnote{Though at first sight such a process 
is $\mathcal{O}(\alpha\,\alpha_s^2)$, the fragmentation function of the outgoing parton into a photon scales as 
$\alpha/\alpha_s$ and the corresponding cross section is of the same order as the direct process~\cite{ACGG}.},
e.g. $q q \to q q\to \gamma\,X$. 
The photons produced in the two first point-like processes  are called ``direct'', the latter ``fragmentation'' photons. 
The Compton channel is particularly interesting as it provides direct information on the proton gluon distribution, 
$g(x,Q^2)$, which is otherwise only {\it indirectly} constrained via the  derivative of the 
proton structure function $\partial F_2(x,Q^2)/\partial \log Q^2$ (``scaling violations'')
in deep-inelastic scattering (DIS) $e$-$p$ collisions~\cite{hera}.\\

%In this context, the study of high-$E_{T}$ prompt photon production 
%the main motivation for the study of high-$E_{T}$ prompt photon production 
%in $pp$ or $p\bar{p}$ collisions is the possibility to constrain the proton gluon distribution, $g(x,Q^2)$, due to the 
%in $pp$ or $p\bar{p}$ collisions provides the possibility to constrain the proton gluon distribution, $g(x,Q^2)$, due to the 
%presence of  the quark-gluon Compton subprocess $ qg\rightarrow \gamma q$ at leading order (LO) in the electromagnetic and 
%strong coupling constants, $\mathcal{O}(\alpha\,\alpha_s)$~\cite{aurenche,vogel}. Although at first sight, the photon 
%is a ``cleaner'' PDF probe compared to hadronic final-states, its theoretical sensitivity %of prompt $\gamma$ data 
%to $g(x,Q^2)$ is partially ``washed-out'' due to the presence of additional LO production mechanisms 
%the main mechanisms of prompt photon production are quark-gluon Compton scattering ($q\,g\to\gamma\,q$) 
%such as quark-antiquark annihilation ($q\,\bar q\to\gamma\,g$) and photons originating 
%from the collinear fragmentation of a final-state parton.\\

%establish a baseline to study medium effects in heavy ion collisions.

Experimentally -- see e.g.~\cite{Vogelsang:1997cq,Aurenche:2006vj} for data compilations %\footnote{The two 
%latest photon measurements of PHENIX~\cite{phenix} and CDF~\cite{cdf} are not included in these data collections.} 
-- many measurements of high-$E_{_T}$ photon production have been carried out in the last 30 years in 
proton-proton (\pp) and proton-antiproton (\ppbar) collisions at fixed-target and collider energies 
spanning center-of-mass (c.m.) energies 
of $\sqrts\approx$~20~--~2000~GeV. 
Apart from results at fixed-target energies from the E706 experiment~\cite{E706}, the agreement between data 
and theory is very good over nine orders of magnitude in the cross sections~\cite{Aurenche:2006vj}. 
In order to identify the prompt photon signal out of the overwhelming background of photons from the %electromagnetic 
decays of $\pi^0$ and $\eta$ mesons produced in the fragmentation of jets, one often requires 
the photon candidates to be isolated from any hadronic activity within a given distance around its
direction. The corresponding measurements are then dubbed ``isolated photons''.
%Prompt photon production in hadronic collisions was discovered at CERN ISR in the late 70s~\cite{?}. 
The most recent\footnote{Not included in the aforementioned data compilations.}
$E_{_T}$-differential spectra for isolated $\gamma$ -- measured at RHIC 
(\pp\ at $\sqrts$~=~200~GeV)~\cite{phenix} and Tevatron~(\ppbar\ at $\sqrts$~=~1.96~TeV)~\cite{d0,cdf} -- 
can be well reproduced by theoretical predictions based on Next-to-Leading Order (NLO) pQCD calculations.\\

Despite the existence of a few hundred data points, the inclusive and isolated photon data measured in \pp\ and \ppbar\
collisions have not been considered in the last ten years in the global-fit analyses performed to determine 
the proton PDFs. The last PDF parametrization to use the photon data was MRST99~\cite{mrst99}. The theoretical 
difficulties to reproduce the fixed-target E706 results at $\sqrts\approx$~30~GeV~\cite{E706} 
motivated their removal from the PDF studies. 
Instead, Tevatron jet data were preferred over photon data in global analysis to constrain the (high-$x$) gluon PDF.
%The determination of the proton PDFs is based on global fit analysis of $F_2$ data from DIS, jets and Drell-Yan.
%In addition, it turns out that the comparisons
The available comparisons of the latest measured isolated photon spectra to NLO have been carried 
out with just one or two PDF parametrizations, e.g. CTEQ6.1M~\cite{cteq6.1} and MRST04~\cite{mrst04}, obtained 
a few years ago. Recently, new PDF sets have become available\footnote{During the completion of this work, 
the NNPDF collaboration released new PDF sets (NNPDF2.0)~\cite{nnpdf2}. Although the gluon PDF has changed, 
it is consistent with the NNPDF1.2 version used here, and we do not expect significant differences 
on the photon production predictions.}% when using one set or another.} 
-- CTEQ6.6~\cite{cteq6.6}, MSTW08~\cite{mstw08} 
and NNPDF1.2~\cite{nnpdf12} -- which include the most up-to-date fits to data from deep-inelastic lepton-proton 
scattering and hadronic collisions as well as various theoretical improvements. The purpose of this paper is twofold. 
Firstly, we want to compare the existing Tevatron data to the theoretical predictions obtained using 
these latest PDF sets. Secondly, we want to study the expected sensitivity of near-future isolated 
photon spectra at the LHC to the same three sets of PDF parametrizations.\\

%including their associated uncertainties.
%In the present article we present a comparison of Tevatron isolated photon data from D0~\cite{d0} and 
%CDF~\cite{cdf} with the NLO QCD calculations implemented in the Monte Carlo (MC) programme \jetphox\. 
%%with varying recent PDF parametrizations.
The paper is organized as follows. In Section~\ref{sec:th} we succinctly remind the theoretical framework 
of our study and present the next-to-leading-order pQCD Monte Carlo (MC) programme \jetphox\ that we have used. 
In Section~\ref{sec:data_th}, we present a comparison of the existing Tevatron isolated photon spectra with 
the \jetphox\ results for various PDFs, theoretical scales and FFs. We then present in Section~\ref{sec:lhc}
the predictions for LHC energies, at central rapidities ($y $~=~0) accessible to the ALICE~\cite{alice}, ATLAS~\cite{atlas} 
and CMS~\cite{cms} experiments as well as at the forward rapidities (up to $y$~=~5) covered by LHCb~\cite{lhcb}. 
We discuss the associated uncertainties in the spectra and the sensitivity to the underlying PDFs. 
We summarize our main findings in Section~\ref{sec:concl}.

%%%%%%%%%%%%%%%%%%%%%%%%%%%%%%%%%%%%%%%%%%
\section{Theoretical framework}
\label{sec:th}

\subsection{Inclusive prompt photon production}
\label{sec:gamma_prompt}

As aforementioned, two types of processes contribute to the prompt photon production cross section 
in \pp\ and \ppbar\ collisions: the `direct' contribution, where the photon is emitted directly 
from a pointlike coupling to the hard parton-parton vertex, and the `fragmentation' 
(sometimes also called `anomalous') contribution, in which the photon originates from the 
collinear fragmentation of a final-state parton. Schematically, the differential photon 
cross section in transverse energy $E_{_T}$ and rapidity $y$ can be written as~\cite{Aurenche:2006vj}: 
%\begin{equation} 
%d\sigma^{\gamma}  =d\sigma^{{dir}}(\mu_{_R},\mu_{_F},\mu_{_{ff}}) +\sum_{k=q,\bar{q},g}d\sigma^{{frag}}_{k}(\mu_{_R},\mu_{_F},\mu_{_{ff}}) \otimes D_{\gamma/k}(\mu_{_{ff}}) 
%\label{eq:1}
%\end{equation}
\begin{eqnarray}
\lefteqn{d\sigma  \equiv d\sigma_{_{dir}}+d\sigma_{_{frag}} =\sum_{a,b=q,\bar{q},g}\int dx_a dx_b \; F_a(x_a;\mu_{_F}^2) F_b(x_b;\mu_{_F}^2) \; \times } \\
& & \left[ d\hat{\sigma}_{ab}^{\gamma}(p_{\gamma}, x_a,x_b; \mu_{_R},\mu_{_F},\mu_{_{ff}})+
\! \sum_{c=q,\bar{q},g}\int^1_{z_{min}}\! \frac{dz}{z^2} d\hat{\sigma}_{ab}^c (p_{\gamma},x_a,x_b,z;\mu_{_R},\mu_{_F},\mu_{_{ff}}) D^{\gamma}_c (z;\mu_{_{ff}}^2)\right] \nonumber
\label{eq:dsigma_nlo}
\end{eqnarray}
where $F_{a}(x_{a},\mu_{_F})$ is the parton distribution function of  parton
species $a$ inside the incoming hadrons $h$ at momentum fraction $x_a$;
$D_{\gamma/k}(z,\mu_{_{ff}})$ is the fragmentation function of parton 
$k$ to a photon carrying a fraction $z$ of the parent parton energy
(integrated from $z_{min}=x_T \cosh y $, with $x_T = 2E_{_T}/\sqrt{s}$, to 1);
%and factorisation scale $\mu_{_F}$; $\alpha_{s}(\mu_{_R})$ is the strong coupling defined
%in the $\overline{\mbox{MS}}$ renormalisation scheme at the  renomalisation scale $\mu_{_R}$.  ,
%where $d\sigma^{{frag}}_{k}$ describes the production of a parton $k$ in a
%hard collision 
and the arbitrary parameters $\mu_{_R}$, $\mu_{_F}$ and $\mu_{_{ff}}$
are respectively the renormalisation, initial-state factorisation, and fragmentation scales
which, loosely speaking, encode any residual dependence of the cross sections to higher-order 
contributions missing in the calculation.
%The probability that a high $E_{_T}$ parton of species $k$ (quark or gluon) ends 
%up splitting into a photon is encoded into a fragmentation function of parton 
%$k$ to a photon, $D_{\gamma/k}(z,\mu_{_{ff}},)$, obtained from fits of the experimental
%$e^+e- \to q \bar q \to \gamma\,X$ data~\cite{ffs}, %defined in some arbitrary fragmentation scheme, 
%at some arbitrary fragmentation scale $\mu_{_{ff}}$.\\
The study provided in this article relies on the calculation of both $d\sigma_{_{dir}}$ and
$d\sigma_{_{frag}}$ at next-to-leading order (NLO) accuracy~\cite{jetphox} in the 
strong coupling $\alpha_{s}(\mu_{_R})$, i.e. all diagrams up to the order $\mathcal{O}(\alpha\alpha_s^2)$ 
are included, defined in the $\overline{\mbox{MS}}$ renormalisation scheme.
The results of the NLO calculation of $d\sigma_{_{dir}}$ have been known for a long time 
\cite{ABDFS}. The calculation of the NLO corrections to $d\sigma_{_{frag}}$ became also
available later~\cite{ACGG,GV,jetphox}. We note that the distinction between $d\sigma_{_{dir}}$ 
and $d\sigma_{_{frag}}$ is arbitrary (only its sum is physically observable): a typical 
case is a bremsstrahlung process from a final-state quark which, depending on the scale, 
can be considered as ``fragmentation'' or as ``NLO direct''.\\
%The knowledge of  $\LMS$, e.g. from deep-inelastic scattering
%experiments, completely specifies  the NLO expression of the running coupling
%$\alpha_{s}(\mu_{_R})$. The NLO correction terms to $d\sigma_{_{dir}}$ and {frag},
%$\kd_{ij}$~\cite{ABDFS,GV} and $\kf_{ij,k}$~\cite{ACGG} respectively, are known
%and their expressions in the $\overline{\mbox{MS}}$ scheme will be used. 
%The dependence of these functions on the kinematical variables $x_1, x_2, z, \sqrt{s}, E_{_T}$ and $y$ has not been explicitly displayed. 
%The structure and fragmentation functions have been determined at the required level of accuracy by NLO fits to the data.
%%%% All the quantities entering the above
%%%% equations have been either calculated ($\kd_{ij}$ and $\kf_{ij,k}$) or

More recently, expressions involving the resummation -- at next-to-leading (NLL) or even 
next-to-next-to-leading (NNLL) logarithmic accuracy -- of threshold and recoil contributions 
due to soft gluon emission, which are %logarithmically 
large close to the phase space boundary where the $E_{_T}$ of the photon is about 
half of the center-of-mass energy ($x_{T} = 2 E_{_T}/\sqrt{s} \to 1)$ have been
%obtained, first in $d\sigma_{_{dir}}$~\cite{los,cmn,cmnov,ko,sv}, and more recently in 
%$d\sigma_{_{frag}}$ as well \cite{deFlorian:2005wf}. This resummation is performed 
obtained for $d\sigma_{_{dir}}$ and $d\sigma_{_{frag}}$~\cite{nll}. 
The effect of this resummation is important at very large\footnote{Note that at the
highest momenta probed, there are also additional corrections due to electroweak boson exchanges, 
which {\it decrease} the photon yields by about 10--20\% within $\ETg\approx$~1~--~2~TeV 
at the LHC~\cite{Kuhn:2005gv}.} photon $E_{_T}$ extending down to values of $x_{T} \approx$~$10^{-1}$, 
%Interestingly, the N(N)LL results 
and provide a much reduced $\mu_{_R}$ and $\mu_{_F}$ scale dependence than the NLO approximation. 
%but, in general, they are consistent with the values of the 
%cross sections obtained at NLO when the scales are reduced somewhat compared
%to the typical $\mu=\ETg$ value~\cite{Aurenche:2006vj}. 
In any case, since we are interested in the low-$x$ region of the PDFs, 
we do not consider such effects, which are not implemented in \jetphox, in this study.\\

%The NLO results roughly agree with the resummed
%calculation in the region populated by fixed target data, when $\mu_{_R}$ and
%$M$ are chosen  $\sim E_{_T}/2$ in the former. 
%Even more recently, a joint summation of both threshold and recoil effects due
%to soft multigluon emission has been performed in \cite{Sterman:2004yk}. 
%Recoil effects are logarithmically enhanced order by order in the $\alpha_{s}$ expansion of
%the $q_{T}$ distribution of a pair $\gamma$-jet; however this logarithmic
%enhancement is washed out by the integration over the jet when passing to the
%single photon inclusive $E_{_T}$ distribution~\cite{aurenche}.

\begin{figure}[htbp]
\centering
\includegraphics[width=9.cm]{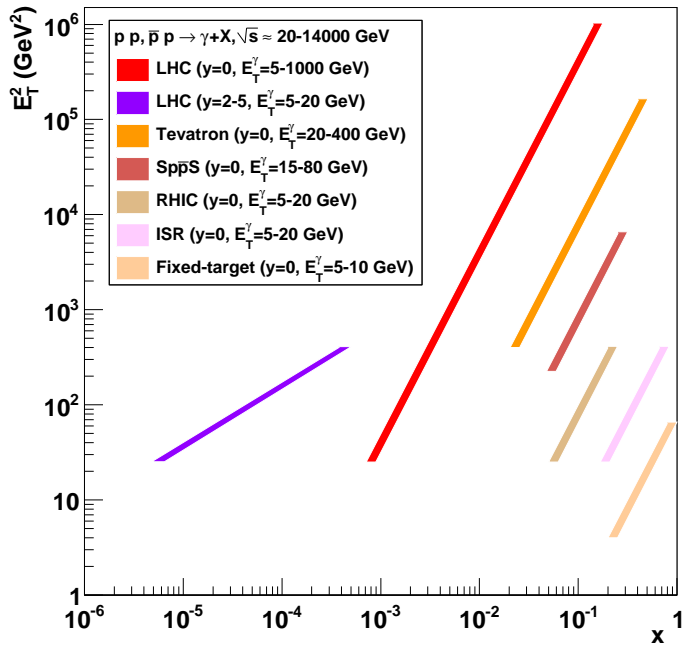}
\caption{Kinematical region probed by existing prompt photon measurements at fixed-target 
(Fermilab) and collider (ISR, RHIC, Sp$\bar{p}$S, Tevatron) energies, and expected range probed 
at the LHC at central ($y$~=~0) and forward ($y$~=~2~--~5) rapidities.} 
\label{fig:kinem}
\end{figure}

In Figure~\ref{fig:kinem} we show the kinematical region covered by the photon measurements in the 
$(x_T, \ E_{_T}^2)$ plane equivalent to the DIS $(x$,$Q^2)$ plane for hard scattering hadronic processes.
The quoted $x$ values are typical $x_T$ values for measurements at midrapidity and $x\approx x_T\cdot\exp($-$y)$
%$x_{min}=x_T\,\exp(-y)/[2-x_T\,\exp(y)]$
at forward rapidities. At midrapidities, photon production at the LHC will probe values 20 times smaller
than at Tevatron because of the factor $\sim$7 increase in the c.m. energy and the lower $E_{_T}$ values 
reachable (e.g. down to $E_{_T} \approx$~5~GeV at ALICE). At the forward rapidities and %moderately low
low $E_{_T}\approx$~5~GeV covered e.g. by LHCb, one can probe the gluon distribution down to very small 
$x\approx$~10$^{-5}$. The combined photon data at all energies -- about 350 data-points collected so far
plus $\cO{100}$ new data-points expected from the four experiments at the LHC -- provide thus constraints 
of the PDFs in a wide ($x,Q^2$) range.

\begin{figure}[htbp]
%\centering
\hspace{-0.9cm}\includegraphics[width=5.7cm,height=6.cm]{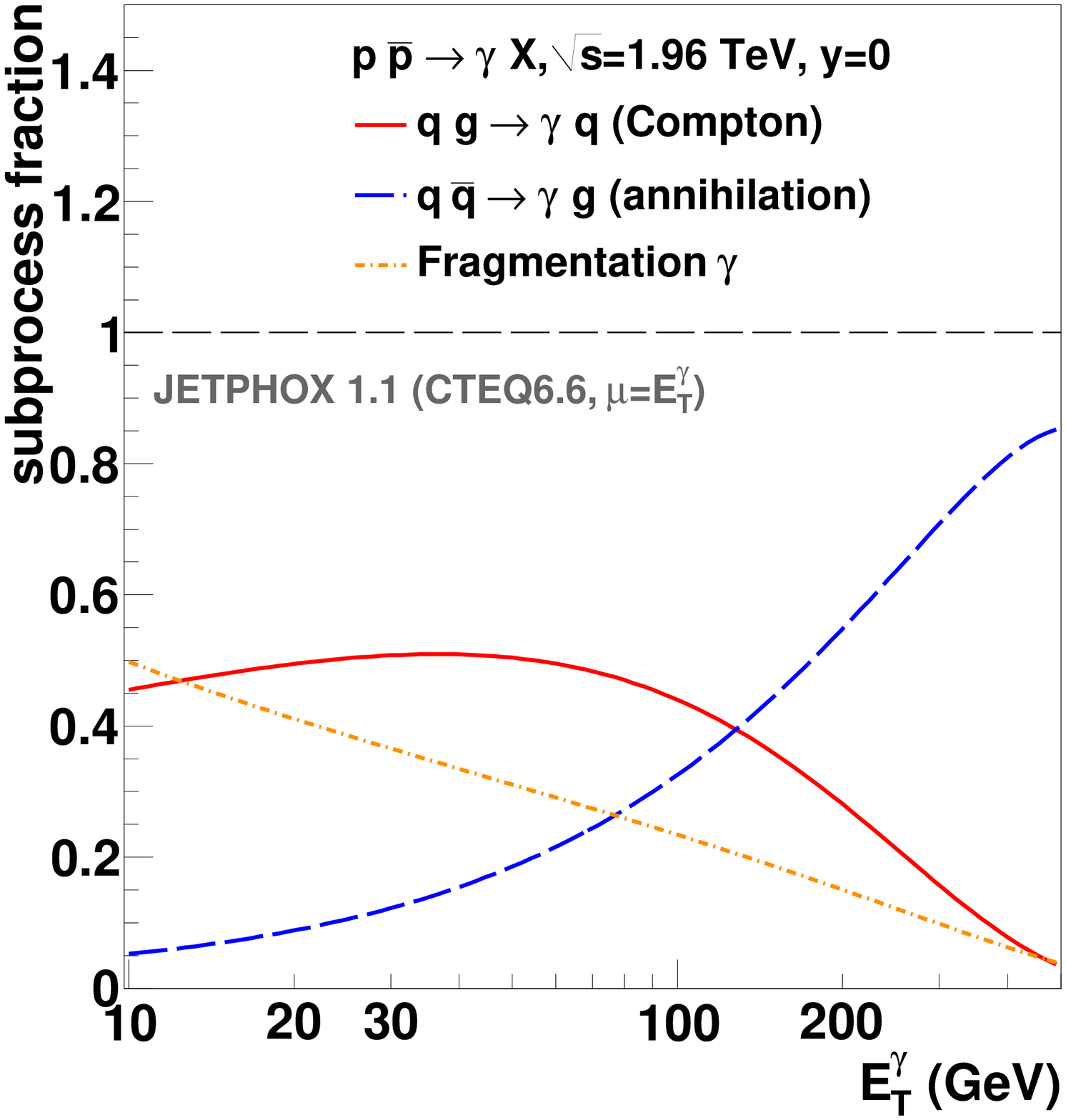} %photon_subprocess_fraction_tevatron.eps}
%\hspace{-0.5cm}\includegraphics[width=6.cm,height=6.cm,viewport =23 0 567 469, clip]{subproc_ppgamma_lhc_inc.eps}
%\hspace{-0.2cm}\includegraphics[width=6.cm,height=6.cm,viewport =23 0 567 469, clip]{subproc_ppgamma_lhcb_inc.eps}
\hspace{-0.2cm}\includegraphics*[width=6.1cm,height=6.cm]{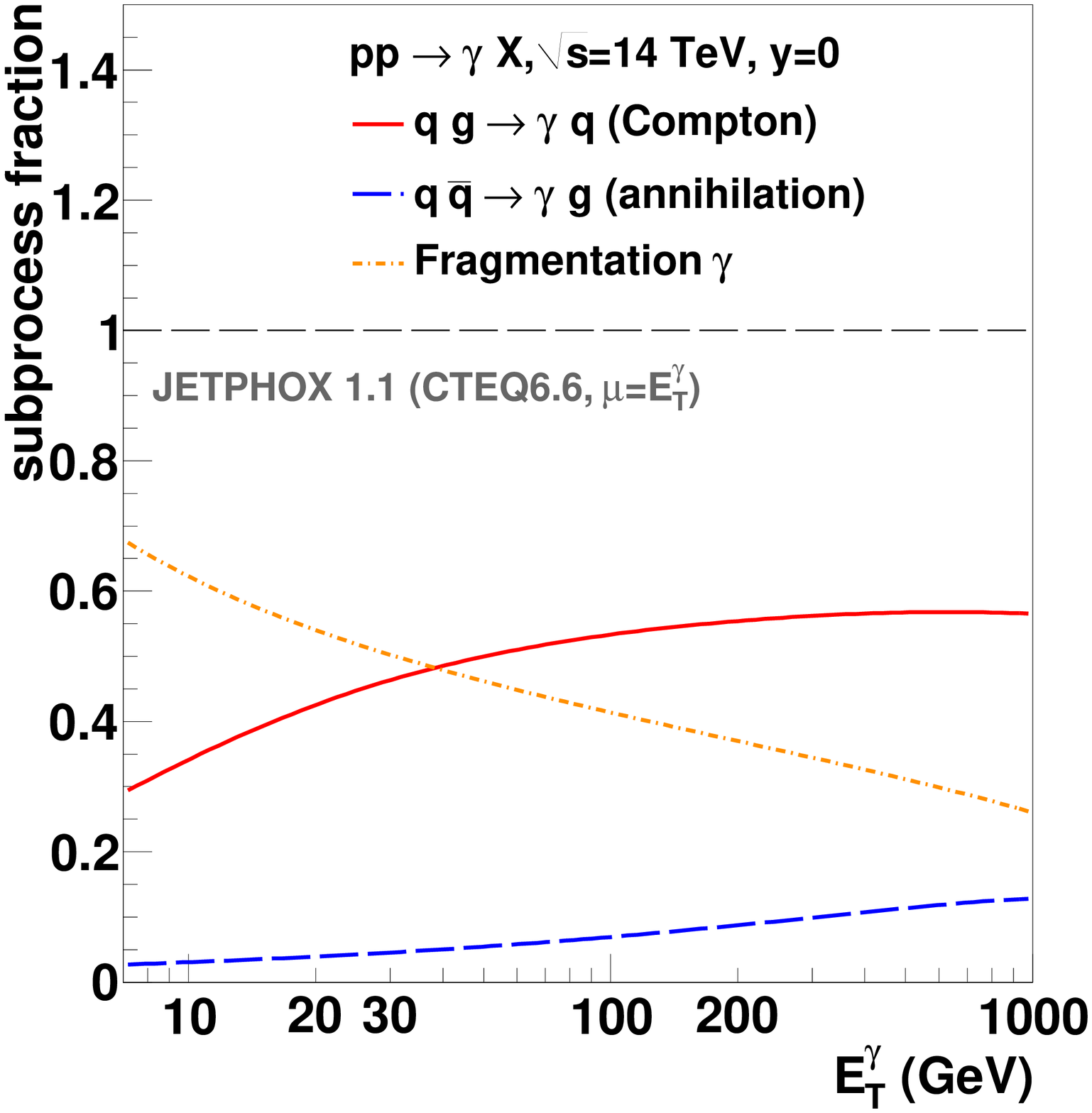}
\hspace{-0.2cm}\includegraphics*[width=5.5cm,height=6.cm]{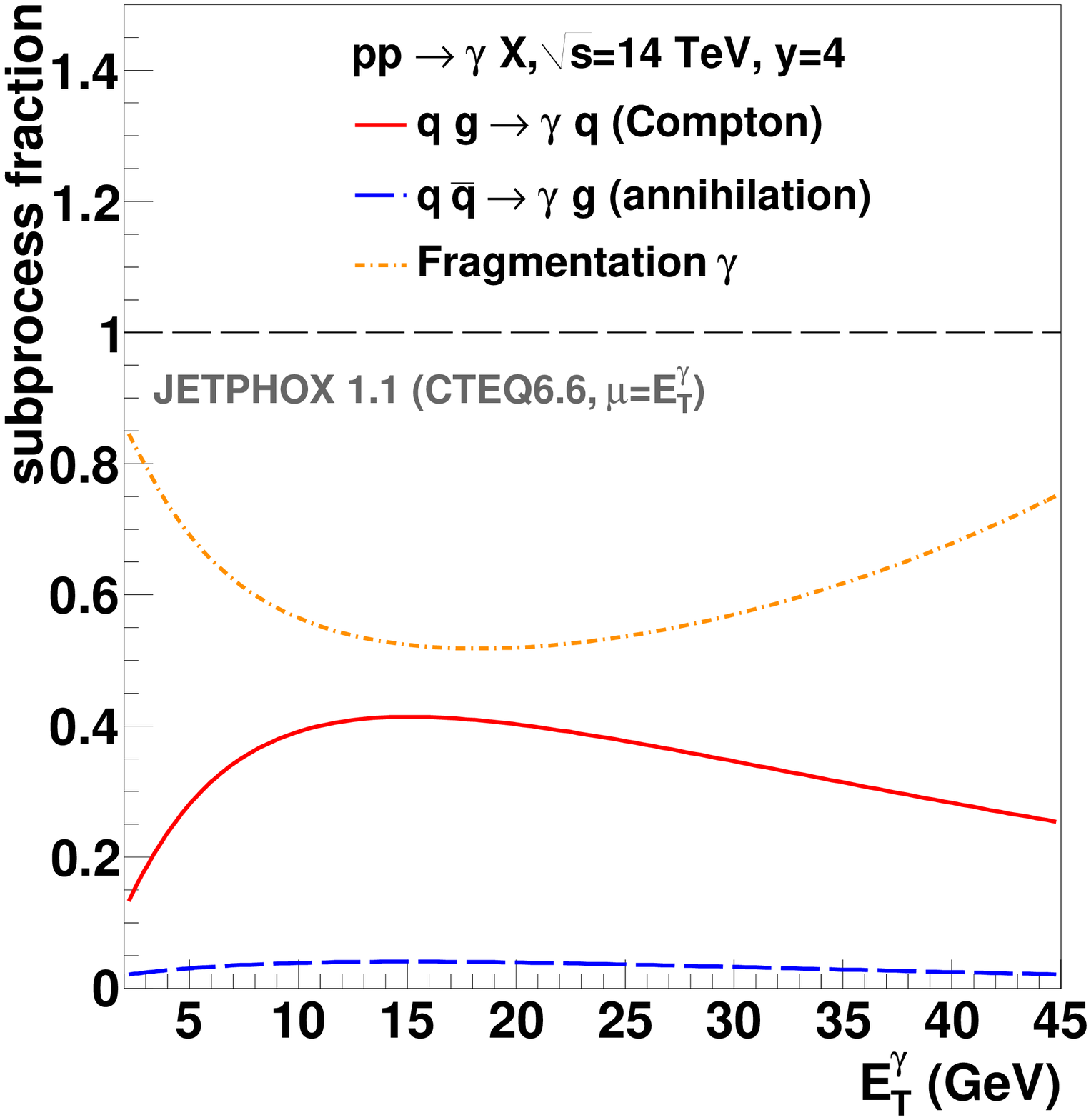}
\caption{Relative contributions of the quark-gluon Compton, \qqbar\ annihilation and fragmentation subprocesses in 
NLO prompt photon production at the Tevatron (left), LHC midrapidity (center) and LHC forward rapidity (right) 
obtained with \jetphox\ (CTEQ6.6 PDF, $\mu$~=~$\ETg$, and BFG-II FFs).}
\label{fig:subproc}
\end{figure}

In Figure~\ref{fig:subproc} we show the relative contributions of each one of the three subprocesses (Compton, 
annihilation and fragmentation) to prompt photon production at Tevatron ($y$~=~0) and LHC ($y$~=~0, and $y$~=~4) 
as a function of the photon $E_{_T}$. [Again, the plots are to be taken indicatively as the weights of the different 
components are scale-dependent.] They have been obtained selecting the corresponding Feynman diagrams at NLO 
with the \jetphox\ MC, setting all scales to $\mu$~=~$\ETg$, and using the CTEQ6.6 parton densities 
and the BFG-II parton-to-photon FFs for the fragmentation photons.
%{\bf FA: Essential to vary mu since the distinction between processes is highly scale-dependent.}
In the low $E_{_T}$ region of the mid-rapidity spectra (below 15~GeV at Tevatron, and 40~GeV at the LHC) 
the fragmentation component dominates the cross sections. At the Fermilab collider, quark-gluon scattering 
is the dominant component up to $E_{_T}\approx$~120~GeV, beyond which %that value of $E_{_T}$ 
the annihilation of valence (anti)quarks from the (anti)proton beams play a preeminent role. At the LHC, 
the Compton process dominates for all $E_{_T}$'s above 45~GeV at $y$~=~0. At forward rapidities at the LHC, 
between 50\% to 80\% of inclusive prompt photons are produced from the collinear fragmentation of a parton 
for all relevant $E_{_T}$'s.\\ % at rapidity $y$~=~4.\\

\subsection{Isolated photon production}
\label{sec:gamma_isol}

In order to strongly suppress the overwhelming background of secondary photons coming from 
the electromagnetic hadron decays (mainly $\pi^{0}$ and $\eta$ mesons) produced inside partonic jets, 
collider experiments require often an isolation criterion on the photon candidates. A standard isolation 
requirement is that within a cone about the direction of the photon defined in rapidity $y$ and 
azimuthal angle $\phi$ by
\begin{equation}
 R = \sqrt{\left( y - y_{\gamma} \right) ^{2} + \left( \phi - \phi_{\gamma} \right) ^{2}}\;,
\label{eq:iso1}
\end{equation}
the accompanying hadronic transverse energy $E_{_T}^{had}$ is less than a given value
\begin{equation}
E_{_T}^{had} \leq E_{_T}^{max} \;. %= \varepsilon_{h}\,\ETg \;.
\label{eq:iso2}
\end{equation}
$R$ is usually taken between 0.4~--~0.7 and $E_{_T}^{max}$ is specified either as a fixed value, 
or as a fixed fraction $\varepsilon_{h}$ (e.g. often 10\%) of the photon $E_{_T}$. Cross sections 
for producing isolated photons have been proven to fulfill the factorization theorem, and are finite 
to all orders in perturbation theory for non zero $R$ and $E_{_T}^{max}$~\cite{Catani:2002ny}.\\

\begin{figure}[htbp]
%\centering
\hspace{-0.9cm}\includegraphics[width=5.7cm,height=6.cm]{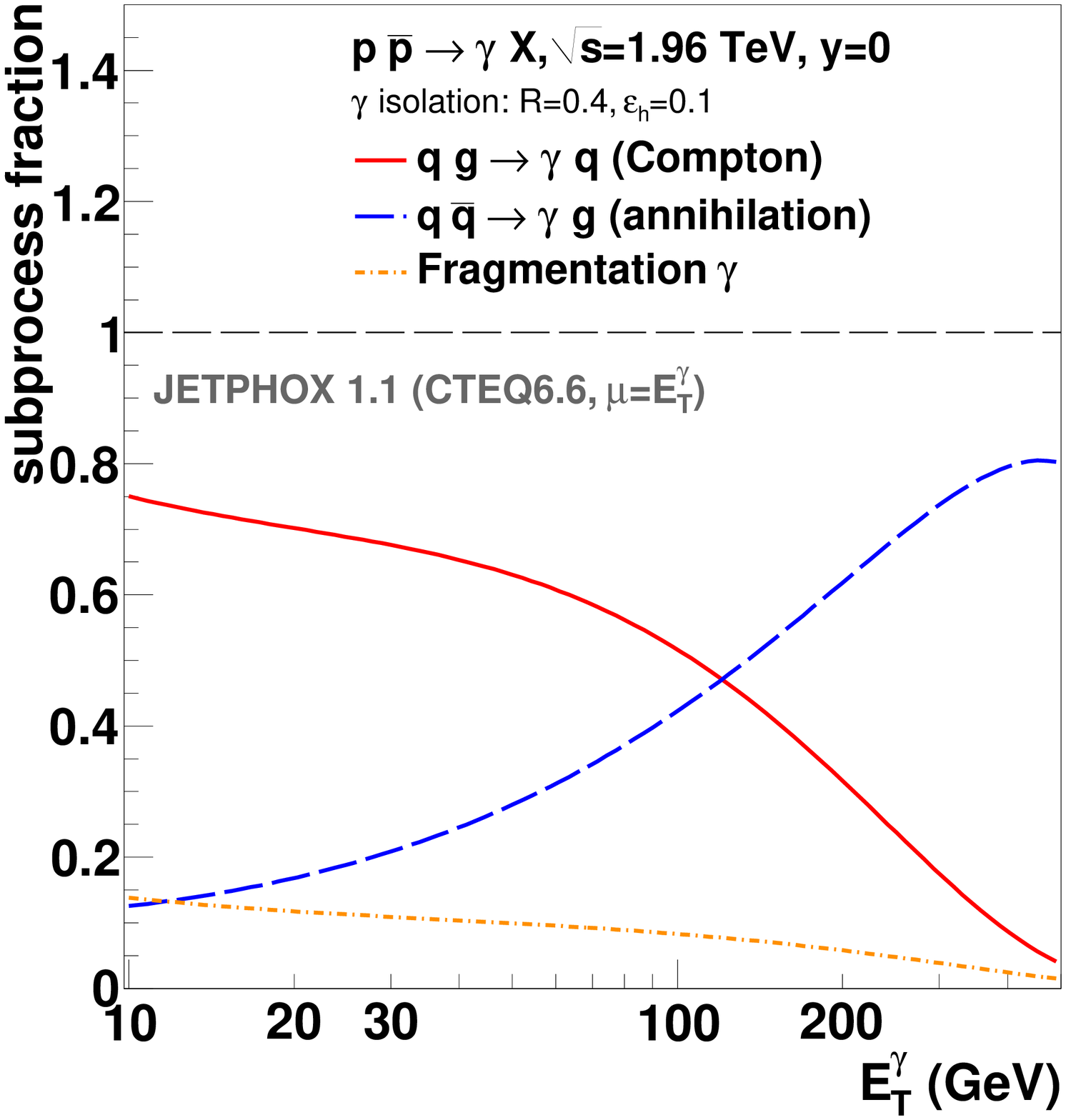} %photon_subprocess_fraction_tevatron.eps}
%\hspace{-0.5cm}\includegraphics[width=6.cm,height=6.cm,viewport =23 0 567 469, clip]{subproc_ppgamma_lhc_inc.eps}
%\hspace{-0.5cm}\includegraphics[width=6.cm,height=6.cm,viewport =23 0 567 469, clip]{subproc_ppgamma_lhcb_inc.eps}
\hspace{-0.2cm}\includegraphics*[width=6.1cm,height=6.cm]{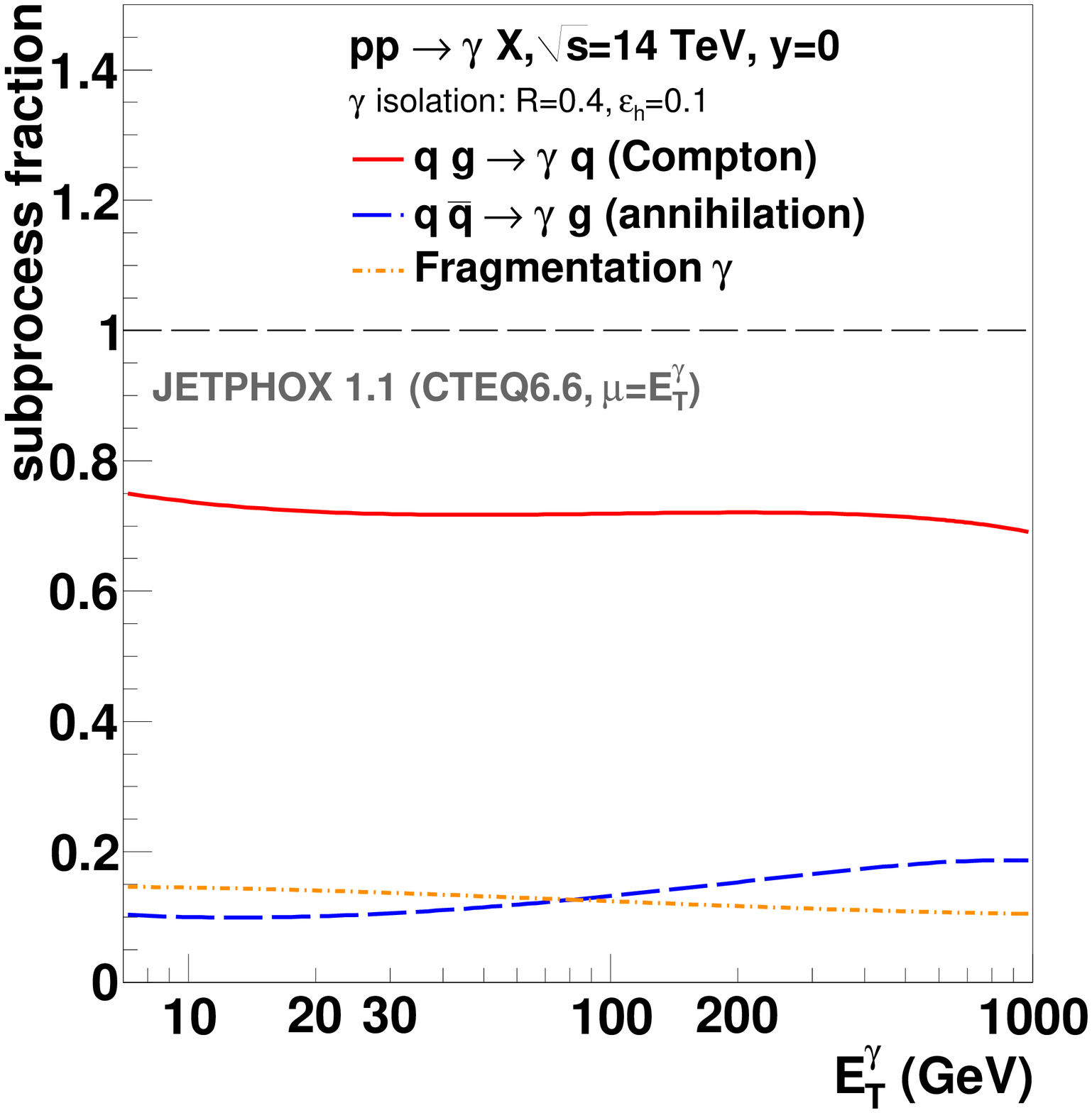}
\hspace{-0.2cm}\includegraphics*[width=5.5cm,height=6.cm]{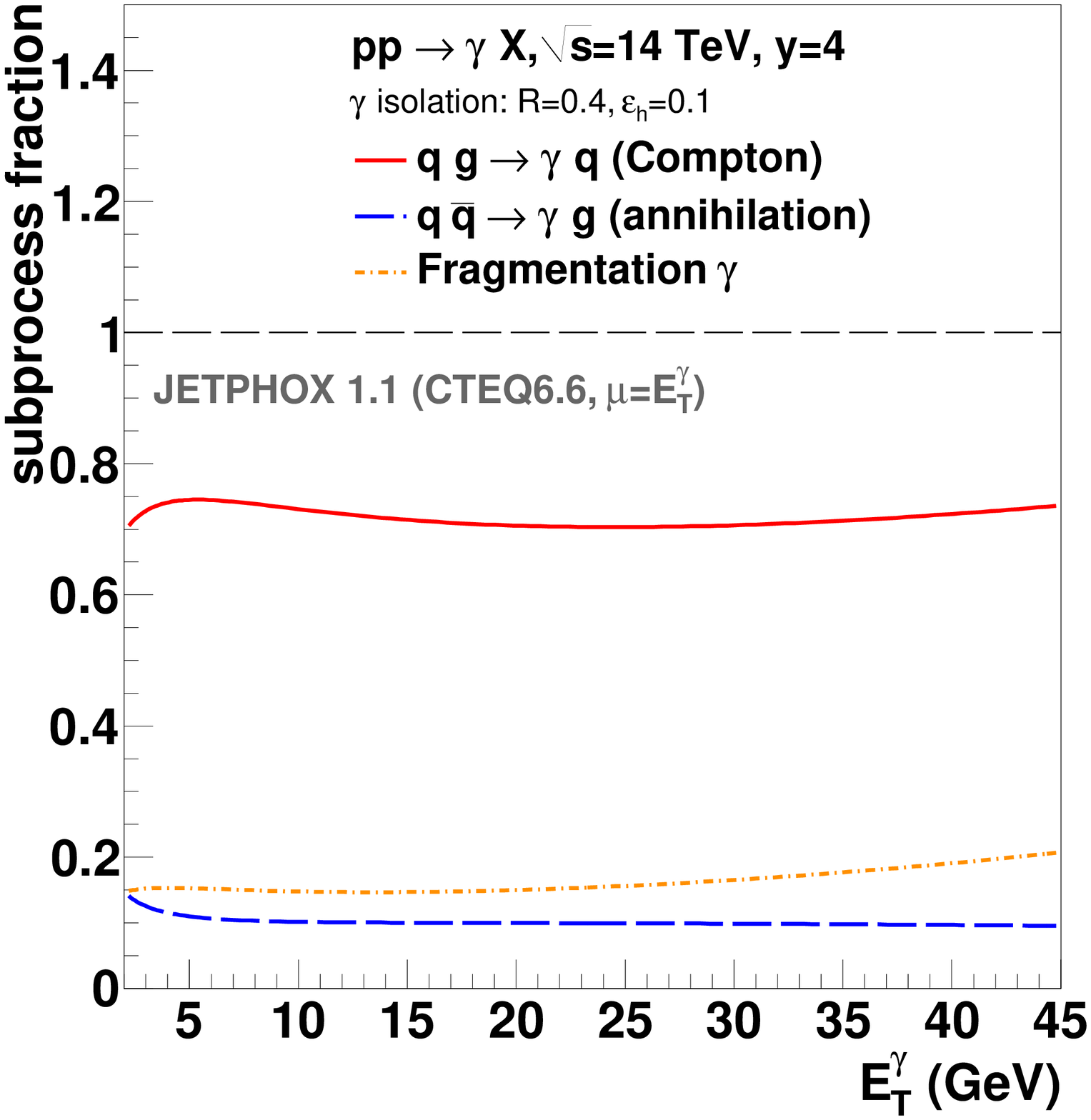}
\caption{Relative contributions of the quark-gluon Compton, \qqbar\ annihilation and fragmentation subprocesses in 
NLO {\it isolated} photon production at the Tevatron (left), LHC midrapidity (center) and LHC forward rapidity (right)
obtained with \jetphox\ (CTEQ6.6 PDF, $\mu$~=~$\ETg$, BFG-II FFs) for an isolation radius $R$~=~0.4 and
a hadron fraction of the photon energy of $\varepsilon_{h}$~=~0.1 inside the cone.}
\label{fig:subproc_isol}
\end{figure}

If one ignores the ``underlying event'' activity in the collision -- due to soft interactions of spectator partons and 
beam remnants~\cite{UE} -- the leading-order \qg-Compton and \qqbar-annihilation photons are emitted in a region completely 
free from hadronic activity (the recoiling quark or gluon is emitted at $\pi$ from it). On the other hand, the fragmentation 
component is usually accompanied by other particles issuing from the hadronization of the same parent parton (jet).
%Whereas the contribution from eq.~(\ref{eq:brem}) amounts\footnote{This statement depends on the choice 
%of scales, especially of $\mu_{_{ff}},$. The order of magnitude given here corresponds to a standard choice 
%$\mu_{_{ff}}, \sim {\cal O}(p_{_T})$.} to roughly a few tens of percent of the contribution from Eq.~(\ref{eq:sig}) 
%at fixed target energies, it becomes dominant at colliders at least in the lower $E_{_T}$ range. 
Yet, the isolated cross section measured experimentally cannot be automatically identified with the direct cross 
section calculated at the Born level. %(i.e. without any contribution of the fragmentation processes). 
First, higher order terms originating in the {\it non-collinear} fragmentation of partons also contribute to 
the isolated cross sections. Second and most important, although isolation cuts such as Eqs. (\ref{eq:iso1}), 
(\ref{eq:iso2}) reduce the $d\sigma_{_{frag}}$ contribution, a fraction of fragmentation photons with 
$z \geq 1/(1 + \varepsilon_{h})$ survive the cuts~\cite{Aurenche:2006vj}. %, which involves the 
%{\it same} fragmentations functions $D_{\gamma/k}(z,\mu_{_{ff}},)$ as in the unisolated case. 
%The dependence on the isolation parameters $R$ and $E_{_T \; max}$ is consistently included in the expression describing the hard subprocess. 
The average $z$-value for fragmentation photons is $\mean{z} \lesssim$~0.6 at the LHC and $\mean{z}\approx$~0.7 
at the Tevatron~\cite{ACGG}, and a typical isolation energy cut of $\varepsilon_{h}$~=~0.1,
corresponding to $1/(1 + \varepsilon_{h}) > 0.9$, suppresses about 60~--~80\% of $d\sigma_{_{frag}}$
(this value is $\ETg$- and scale-dependent). %is suppressed by isolation cuts.
%On the other hand, notice that, at fixed targets, $\mean{z} \sim 0.9$: in practice photons from fragmentation 
%at fixed targets are scarcely accompanied by hadrons, they are {\it de facto} isolated.  
%than $(1 + \varepsilon_{h})^{-1}$, so that $d\sigma_{_{frag}}$ is quite suppressed by isolation cuts.
We see this in more detail in Fig.~\ref{fig:subproc_isol} where we show the subprocesses contributions to the 
{\it isolated} photon cross section. At variance with Fig.~\ref{fig:subproc} for the inclusive prompt-$\gamma$ 
case, we can see that a very significant part of the fragmentation component is suppressed after applying typical 
isolation cuts ($R$~=~0.4, $\varepsilon_h$~=~0.1). The Compton process now clearly dominates the photon yield below 
$\ETg\approx$~120~GeV at Tevatron and accounts for about 3/4 of isolated $\gamma$ production for all $E_{_T}$'s at 
the LHC. These results confirm the interesting potential of isolated photon hadroproduction to constrain the gluon 
density.\\

\subsection{\jetphox\ Monte Carlo}
\label{sec:jetphox}

The present study relies on the implementation of the photon NLO calculation of both $d\sigma_{_{dir}}$ and 
$d\sigma_{_{frag}}$ in the \jetphox\ Monte Carlo (MC) programme~\cite{jetphox,Catani:2002ny}.
The main advantage of the \jetphox\ MC is that one can easily account for any kind of experimental cuts 
(e.g. on kinematics and/or isolation) implementable at the partonic level. In addition, one can match naturally 
the binning of experimental data by histogramming of the partonic configurations generated.
%It is the only available code including 
%both $d\sigma_{_{dir}}$ and $d\sigma_{_{frag}}$ at NLO in a Monte carlo approach.
%Details on the principles and implementation of this code can be found in
%\cite{jetphox,Catani:2002ny}.
All the NLO results provided in the following sections are obtained using\footnote{Note that 
although each PDF set uses a slightly different reference value: $\alpha_s(M_Z) = 0.118$ (CTEQ6.6), 
$\alpha_s(M_Z) = 0.119$ (NNPDF1.2), and $\alpha_s(M_Z) = 0.12018$ (MSTW08), the resulting 
differences in the $\gamma$ cross sections obtained using the slightly different coupling choices 
are very small.} $\alpha_s(M_Z) = 0.118$, 
with up to 5 active quark flavours (the lowest $\ETg$ considered in this work is close to $m_b\approx$~4~\GeVcc). 
We switched off the box diagram $g\,g\to g\gamma$ in the calculations because its contribution 
to the single inclusive spectrum is found to be of just a few percent. The CTEQ6.6, MSTW08 
and NNPDF1.2 PDFs were interfaced to \jetphox\ via the \lhapdf\ (version 5.7.1) package~\cite{lhapdf}.
%and the parton-to-photon fragmentation functions of set BFGW (set II)~\cite{BFG}.
Whenever the scales $\mu_{_R}$, $\mu_{_F}$ and $\mu_{_{ff}},$ are given a common value, the 
latter is noted $\mu$ hereafter.\\ %The $\overline{MS}$ scheme is used throughout. \\

In \jetphox\ the NLO calculations are performed at the parton level %for the QCD hard process 
and do not account for hadronisation effects which can be potentially large at low $E_{_T}$. In addition, they do 
not include the soft hadronic activity from the underlying event (parton spectator collisions, beam remnants ...)
whose transverse energy can also fall inside the isolation cone. However, the experimental data at Tevatron 
have been corrected for such effects. In the case of the CDF analysis, the correction is found to amount about 
9\% of the photon yield, independent of its $E_{_T}$ above 30~GeV~\cite{cdf}. Studies at the LHC using various 
tunes of the underlying event with \pythia~\cite{pythia6.4} indicate that the loss of isolated photons due to 
hadronic activity within the isolation cone is of about 20\% at $\ETg\approx$~5~GeV and negligible above 70~GeV 
(for radius $R$~=~0.4 and hadronic energy fraction  $\varepsilon_h$~=~0.1)~\cite{pythia_jetphox}.

%%%%%%%%%%%%%%%%%%%%%%%%%%%%%%%%%%%%%%%%%%%%
\section{Comparison of Tevatron data to NLO pQCD}
\label{sec:data_th} 

During the Tevatron Run-II, the \dzero\ and CDF collaborations collected respectively 0.32~fb$^{-1}$ and 2.5~fb$^{-1}$ 
worth of photon data in \ppbar\ collisions at $\sqrt{s} = 1.96$ GeV~\cite{d0,cdf}. Such data sample allowed both 
experiments to measure isolated prompt photons in the $E_{_T}$ range from roughly 20~GeV to about 400 GeV. 
The detailed conditions of the measurements are as follows
\begin{itemize}
\item \dzero : Kinematics: $E_{_T}^\gamma$~=~20~--~300~GeV, $|y^\gamma| <$~0.9. % Lumi = 320 pb-1
Isolation criterion: $R=0.4$, $\varepsilon_h = E_{_T}^{had}/E_{_T}^{\gamma}<$~0.1.
\item CDF: Kinematics: $E_{_T}^\gamma$~=~30~--~400~GeV, $|y^\gamma| <$~1.0.  % Lumi = 2.5 fb-1
Isolation criterion: $R=0.4$, $E_{_T}^{had}~<$~2~GeV.
\end{itemize}

We have used the kinematics and isolation cuts in \jetphox\ corresponding to each measurement
and calculated the resulting isolated photon spectrum with the same $E_{_T}$ binning as the experimental spectra. 
The calculations have been repeated for three PDF sets (CTEQ6.6M, MSTW08, NNPDF1.2), three scales 
($\mu$=0.5$\ETg$, $\ETg$, 2$\ETg$) and two photon FFs (BFG-I and BFG-II)~\cite{BFG}. The results 
are presented and discussed in the next subsections.\\
%We impose the \dzero\ isolation criterion by requiring that the hadronic transverse energy measured 
%in a cone of radius $R= \sqrt{\Delta\phi^2 + \Delta y^2} = 0.4$ around the photon is smaller 
%than 10\% of the photon's $E_{_T}$. 

\begin{figure}[htbp]
\centering
\includegraphics[width=7.7cm,height=9.cm]{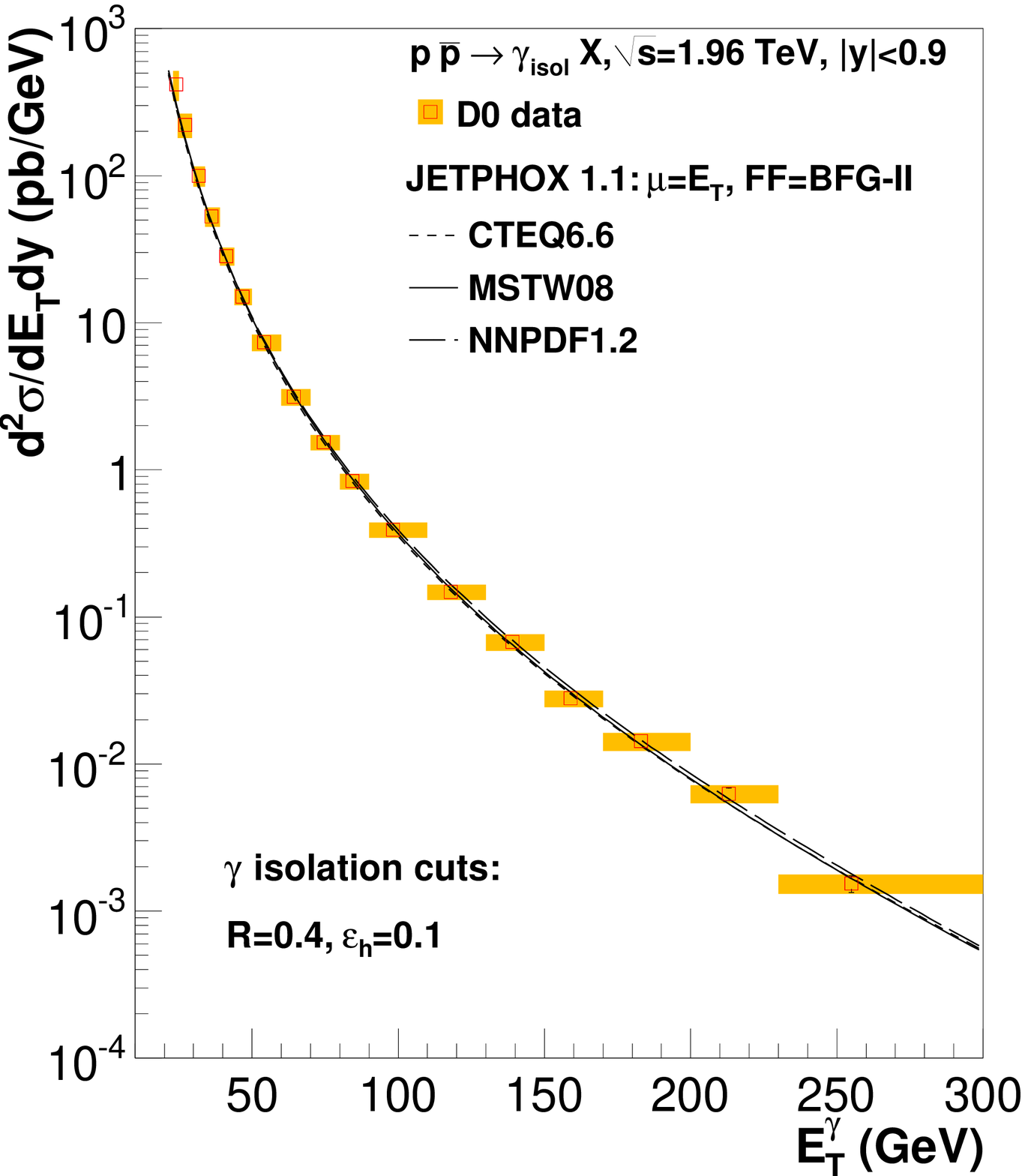}
%ppbar_photons_d0_vs_jetphox_pdfall_mu1.eps}
\includegraphics[width=8.3cm,height=9.cm]{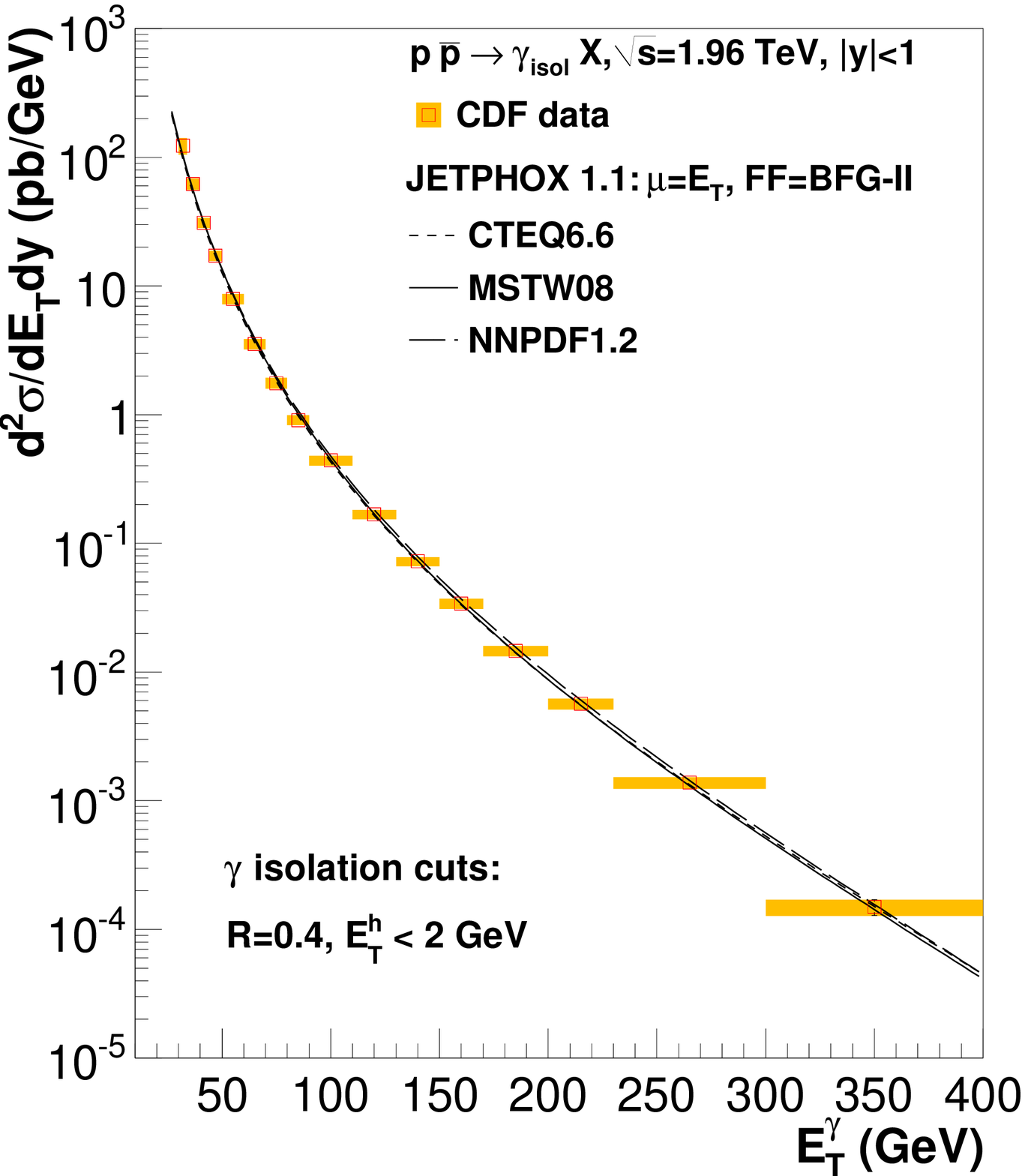}
%ppbar_photons_cdf_vs_jetphox_pdfall_mu1.eps}
\caption{Isolated photon $E_{_T}$-differential cross section at mid-rapidity in \ppbar\ at $\sqrts$~=~1.96~TeV 
measured by \dzero\ (left) and CDF (right) compared to \jetphox\ ($\mu = \ETg$, BFG-II) for three different PDFs: 
CTEQ6.6, MSTW08 and NNPDF1.2. The (orange) bands indicate the experimental systematic errors.}
\label{fig:pdf_dep}
\end{figure}

%%%%%%%%%%%%%%%%%%%%%%%%%%%%%%%%%%%%%%%%%%%%
\subsection{Parton distribution functions dependence}
\label{sec:PDFs}

\begin{figure}[htbp]
\centering
\includegraphics[width=7.7cm,height=6.5cm]{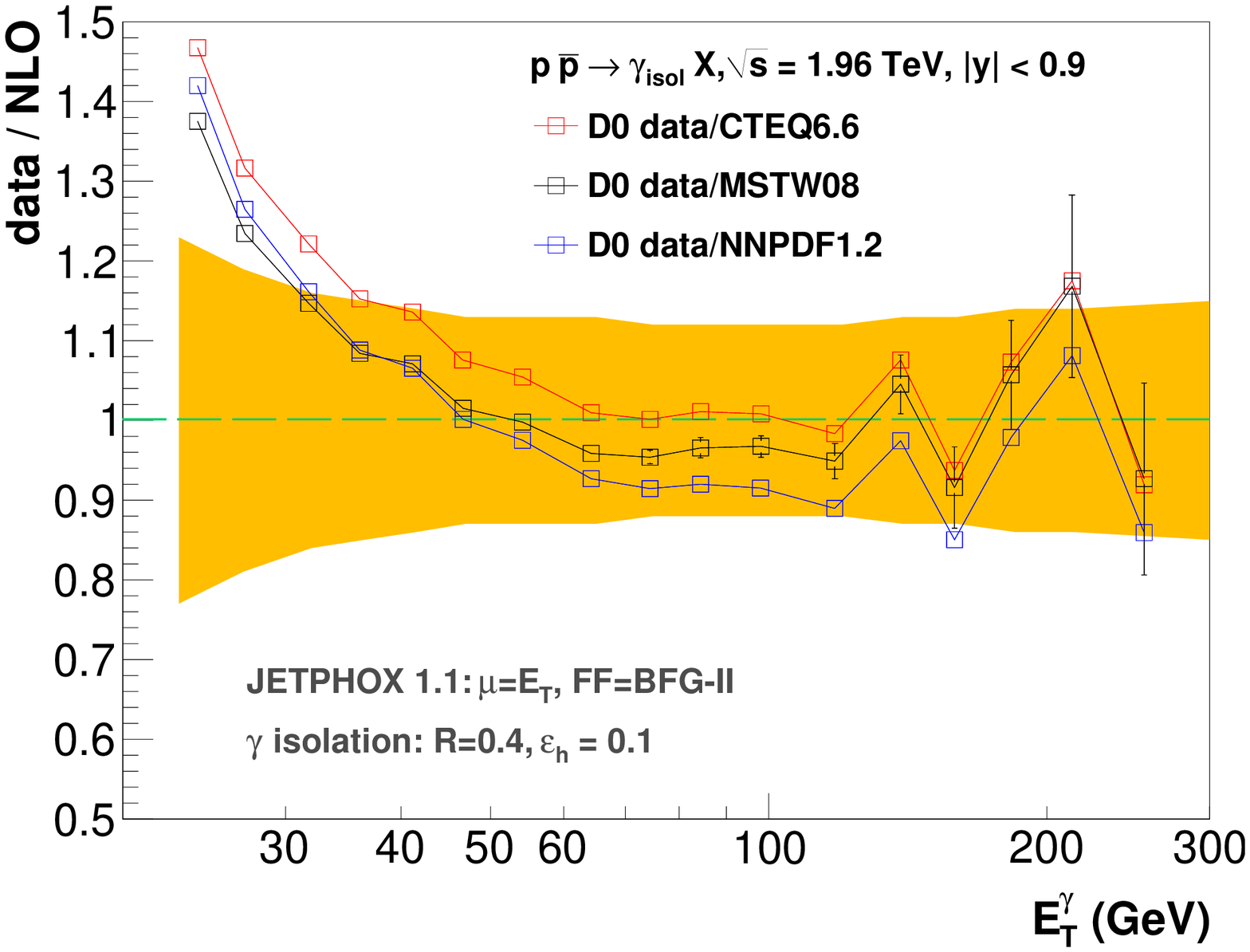}
%fratio_ppbar_photons_d0_vs_jetphox_pdfall_mu1.eps}
\includegraphics[width=8.3cm,height=6.5cm]{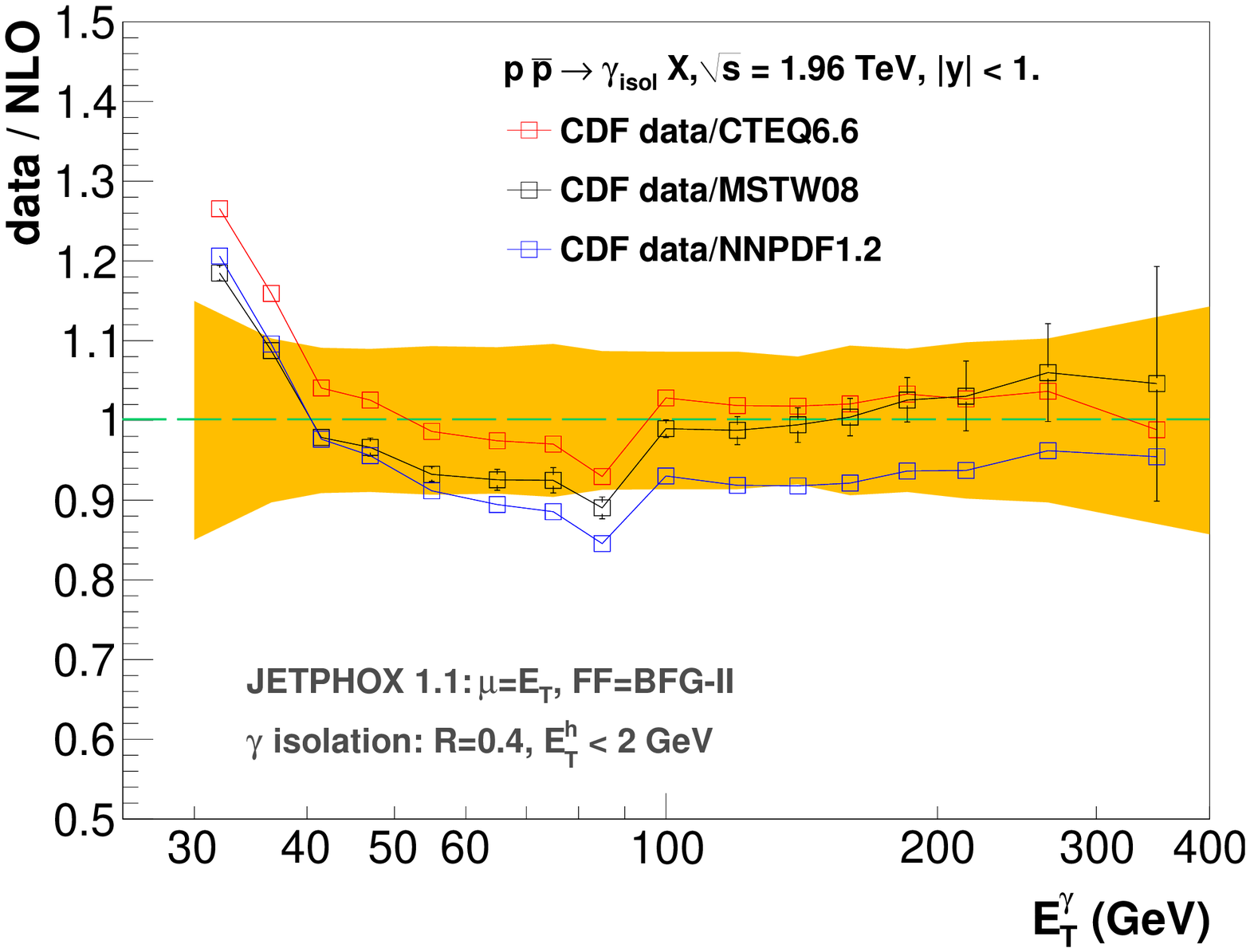}
%fratio_ppbar_photons_cdf_vs_jetphox_pdfall_mu1.eps}
\caption{Ratio of isolated photon %$E_{_T}$-differential cross section at mid-rapidity 
midrapidity spectra in \ppbar\ at $\sqrts$~=~1.96~TeV
measured by \dzero\ (left) and CDF (right) over the \jetphox\ ($\mu = \ETg$, BFG-II) predictions for three PDFs: 
CTEQ6.6, MSTW08 and NNPDF1.2. For clarity, only the MSTW08 ratio is shown with propagated statistical errors
from the data. The lines between data/NLO points are to guide the eye. The (orange) bands indicate 
the systematic uncertainties of the measurements.}
\label{fig:pdf_dep_ratio}
\end{figure}

The \dzero\ and CDF isolated photon spectra are compared  in Fig.~\ref{fig:pdf_dep} to the corresponding 
\jetphox\ predictions for three different PDF parametrizations (CTEQ6.6, MSTW08 and NNPDF1.2).
The theoretical scales have been set to $\mu=\ETg$ and the parton-to-photon FFs to the BFG-II set.
The three theoretical spectra are very similar and they all agree well with the data within the experimental 
uncertainties given by the error bars (resp. bands) representing the statistical (resp. systematical) errors.
%Using MRST 2004 \cite{mrst04} instead of CTEQ6M changes the predictions by $\pm 2$ \%. 
The isolated photon spectrum obtained using MSTW08, for both sets of experimental cuts, is mostly in between 
the spectra obtained with the two other PDFs, except at small-$\ETg$ where MSTW08 results in a somewhat
lower cross section, closer to the data. This is better seen in Fig.~\ref{fig:pdf_dep_ratio} where we
plot the ratio data/theory calculated with the three choices of PDF. On average, the NNPDF1.2 (resp. CTEQ6.6) 
prediction is a few percents higher (resp. lower) than the MSTW08 distribution. As aforementioned, all 
PDFs reproduce well in magnitude and shape the experimental spectrum within the experimental uncertainties, 
dominated by the energy-scale uncertainty (orange bands in the plots).

A simple $\chi^2$ analysis of the full spectrum to the theoretical predictions, accounting for the experimental 
uncertainties (statistical and systematical  added in quadrature) yields $\chi^2/ndf$ below 1 for the 
three PDF predictions. For the default scale-choice shown, $\mu=\ETg$, MSTW08 reproduces a bit 
better both the CDF and \dzero\ results. One has to note, however, that in the $\ETg$ range below 
$\sim$40~GeV a shape difference is apparent, with all NLO curves systematically under-predicting 
the steepness of the cross section by up to around 40\% in the lowest $\ETg$ bin. The source of
such a disagreement, which is about a $2\sigma$ effect in the case of \dzero, is unclear. 
Theoretically, the inclusion of small-$x$ resummation effects~\cite{jrojo2010} does not help to 
improve the data-NLO agreement.% whereas NNPDF1.2 is slightly closer to the \dzero\ measurement.}.

%%%%%%%%%%%%%%%%%%%%%%%%%%%%%%%%%%%%%%%%%%%%
%\subsection{Dependence on the theoretical scales}
\subsection{Scale dependence}
\label{sec:scales}

\begin{figure}[htbp]
\centering
\includegraphics[width=9.cm]{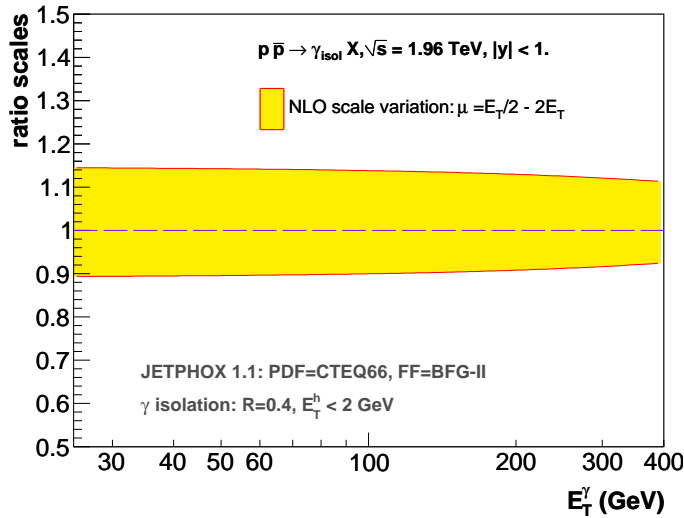}
%\includegraphics[width=8.cm,height=6.cm]{fratio_ppbar_isophot_cdf_vs_jetphox_cteq66_allmu_bfg2.eps}
%fratio_ppbar_photons_cdf_vs_jetphox_pdfcteq66_muall.eps}
\caption{Scale uncertainty of the \jetphox\ NLO predictions for isolated photon production at Tevatron 
(CDF isolation cuts) plotted as the ratio of the spectrum obtained with $\mu = 0.5\ETg,2\ETg$ over that 
with $\mu = \ETg$.}
\label{fig:scale_dep}
\end{figure}

As mentioned in Section~\ref{sec:gamma_prompt}, the truncation of the $\alpha_s$ expansion of the 
pQCD cross section for prompt photon production, 
Eq.~(\ref{eq:dsigma_nlo}), at a given order introduces an arbitrary dependence on three unphysical 
scales: the renormalization scale $\mu_{_R}$ which appears in the evolution of the coupling, the 
factorization scale $\mu_{_F}$ associated to collinear singularities in the PDFs, and the fragmentation 
scale $\mu_{_{ff}}$ related to collinear singularities in the FFs. Since those scales are unphysical, the 
theory can be considered reliable only where the predictions are stable with respect to variations of them.
Usually scale variations of a factor of two around the ``natural'' physical scale of the scattering process,
often taken as the $E_{_T}$ of the photon, are considered: $\mu$~=~$\ETg/2$~--~$2\ETg$. 
%%FA: Why not pTg ? at forward rapidities?
%In Fig.~\ref{fig:scale_dep}, we compare the \dzero\ and CDF spectra to \jetphox\ calculations 
In Fig.~\ref{fig:scale_dep}, we show the ratio of the \jetphox\ isolated $\gamma$ spectra at Tevatron
obtained with scales set to $\mu~=~\ETg/2$ and $\mu~=~2\ETg$ over the same spectrum for $\mu~=~\ETg$.
In all three cases, we use the CTEQ6.6 PDFs\footnote{The scale-dependence obtained with the two other PDF 
parametrizations are similar.} and the BFG-II FFs. %for the scales $\mu$~=~$\ETg/2$, $\ETg$ and~$2\ETg$. 
The sensitivity to the changes in the theoretical scale $\mu$ is of about $\pm 10$\% in the whole 
$E_{_T}$-range. The choice of a scale value $\mu=\ETg$ agrees better with the %central values of the 
experimental spectrum for the MSTW08 and CTEQ6.6 PDFs whereas the NNPDF1.2, whose central spectrum
is a bit higher than the two others, favours a larger $\mu~=~2\ETg$ scale.

%Theory and data are compared in Fig.~\ref{fig:1} and \ref{fig:2}. 
%In Fig.~\ref{fig:1}, all the scales have been set equal to $E_{_T}$. We see that 
%The agreement between data and NLO pQCD is excellent in the whole $E_{_T}$-range in
%which the cross section falls by about six orders of magnitude. 
%The experimental errors are relatively large, of the order of the theoretical
%uncertainty when varying the common scale from  $E_{_T}/2$ to $2E_{_T}$. However, the
%standard choice  $E_{_T}/2$ reproduces the data extremely well over the whole $E_{_T}$
%range in which the cross section varies by  a factor $5\,10^5$.
%The present experimental errors have the size of the variations 
%coming from the scale changes; with this accuracy there is no evidence of any 
%systematic deviation of the theory with respect to data.

%\begin{figure}[htbp]
%\centering
%\includegraphics[width=8.cm,height=8.cm]{ppbar_photons_d0_vs_jetphox_pdfcteq66_muall.eps}
%\includegraphics[width=8.cm,height=8.cm]{ppbar_photons_cdf_vs_jetphox_pdfcteq66_muall.eps}
%\caption{[y-units: pb/GeVc-1] 
%Isolated photon $E_{_T}$-differential cross section at mid-rapidity in \ppbar\ at $\sqrts$~=~1.96~TeV measured by \dzero\ (left)
%and CDF (right) compared to \jetphox\ (CTEQ6.6, BFG-II) for three different scales: $\mu = 0.5\ETg - 2\ETg$.}
%\label{fig:scale_dep}
%\end{figure}

%A more precise comparison is shown in Fig.~\ref{fig:scale_dep_ratio} for the ratio data/theory calculated 
%with three choices of scales between $E_{_T}/2$ and $2E_{_T}$. The sensitivity to 
%the changes in the common scale $\mu$ is of some $\pm 10$ \% in the whole 
%$E_{_T}$-range. Comment on the discrepancy at low pT and the possible ``slope'' in the data/theory ratio ...
%CDF data versus theory for the choice of scales as shown in the figure.

%%%%%%%%%%%%%%%%%%%%%%%%%%%%%%%%%%%%%%%%%%%%
\subsection{Dependence on the parton-to-photon fragmentation functions (FFs)}
\label{sec:FFs}

The production of fragmentation photons in hadronic collisions %to the total prompt $\gamma$ hadroproduction 
is completely encoded in the parton-to-photon FFs, $D_{\gamma/k}(z,\mu_{_{ff}})$ obtained from fits~\cite{BFG,GRV} 
to photon production data at LEP ($e^+e^-\to q\bar{q}(g)\to \gamma X$)~\cite{lep_gamma}. In \jetphox\
two choices of FFs are available: the BFG set I (``small gluon'') and II (``large gluon'')~\cite{BFG}. 
Although both FFs reproduce the LEP data, mostly sensitive to the high $z=p_{\gamma}/p_{parton}$ 
range of the FFs, at small $z$ the gluon-to-photon fragmentation function of set I is significantly lower 
than the one of set II. Otherwise, the quark-to-photon fragmentations are identical in BFG-I and BFG-II.
%leading to a significant difference in the $d\sigma/dz_\gamma$ distribution. 
%In Fig.~\ref{fig:ff_dep} we show the %\dzero\ and CDF isolated spectra to the 
%ratio of theoretical predictions for Tevatron isolated spectra (for the CTEQ6.6 PDF and fixed scales $\mu=\ETg$) 
%obtained with BFG-I over BFG-II. The spectra obtained with both FFs have the same $E_{_T}$ dependence and 
%only absolute differences in the cross sections smaller than 5\% are observed. 
We have run prompt and isolated photon production in the Tevatron kinematic range
with \jetphox\ at NLO with the two FF sets.
The inclusive prompt $\gamma$ spectrum obtained with BFG-I is just a few percent smaller 
than the one obtained with BFG-II, below $\ETg\approx$~80~GeV. The corresponding BFG-I and BFG-II 
isolated photon spectra are virtually identical. This is not unexpected since a significant 
fraction, up to 80\% (see Fig.~\ref{fig:subproc_isol}), of the fragmentation photons are removed 
by standard isolation cuts $R=0.4$, $\varepsilon_h = $~0.1. %~(\ref{eq:iso1})-(\ref{eq:iso2}).
%the experimental data are not sensitive to the choice of FF.

%\begin{figure}[htbp]
%\centering
%\includegraphics[width=8.cm,height=8.cm]{ppbar_photons_d0_vs_jetphox_pdfcteq66_mu1_ffall.eps}
%\includegraphics[width=8.cm,height=8.cm]{ppbar_photons_cdf_vs_jetphox_pdfcteq66_mu1_ffall.eps}
%\caption{[Worth to show ??] Isolated photon $E_{_T}$-differential cross section at mid-rapidity in \ppbar\ at $\sqrts$~=~1.96~TeV 
%measured by \dzero\ (left) and CDF (right) compared to \jetphox\ (CTEQ6.6, $\mu = \ETg$) for two different FFs: BFG-I and BFG-II.}
%\label{fig:ff_dep}
%\end{figure}

%\begin{figure}[htbp]
%\centering
%\includegraphics[width=8.cm,height=6.cm]{fratio_ppbar_photons_d0_vs_jetphox_pdfcteq66_muall.eps}
%\includegraphics[width=8.cm,height=6.cm]{fratio_ppbar_photons_cdf_vs_jetphox_pdfcteq66_muall.eps}
%\caption{Ratio of isolated photon $E_{_T}$-differential cross section at mid-rapidity in \ppbar\ at $\sqrts$~=~1.96~TeV 
%measured by \dzero\ (left) and CDF (right) over the \jetphox\ (CTEQ6.6, $\mu = \ETg$) predictions for two different FFs: BFG-I and BFG-II.}
%\label{fig:ff_dep_ratio}
%\end{figure}

%A more precise comparison is shown in Fig.~\ref{fig:ff_dep_ratio} for the ratio data/theory calculated 
%with two choices of FFs. The sensitivity to the changes in the FF is of some $\pm ??$ \% in the whole $E_{_T}$-range. 

%%%%%%%%%%%%%%%%%%%%%%%%%%%%%%%%%%%%%%%%%%%%
%\clearpage

\section{Isolated photon production at the LHC}% and the proton PDFs}
\label{sec:lhc}

\begin{figure}[htbp]
%\vspace{9pt}
\centering
\includegraphics[width=8.65cm,height=9.2cm]{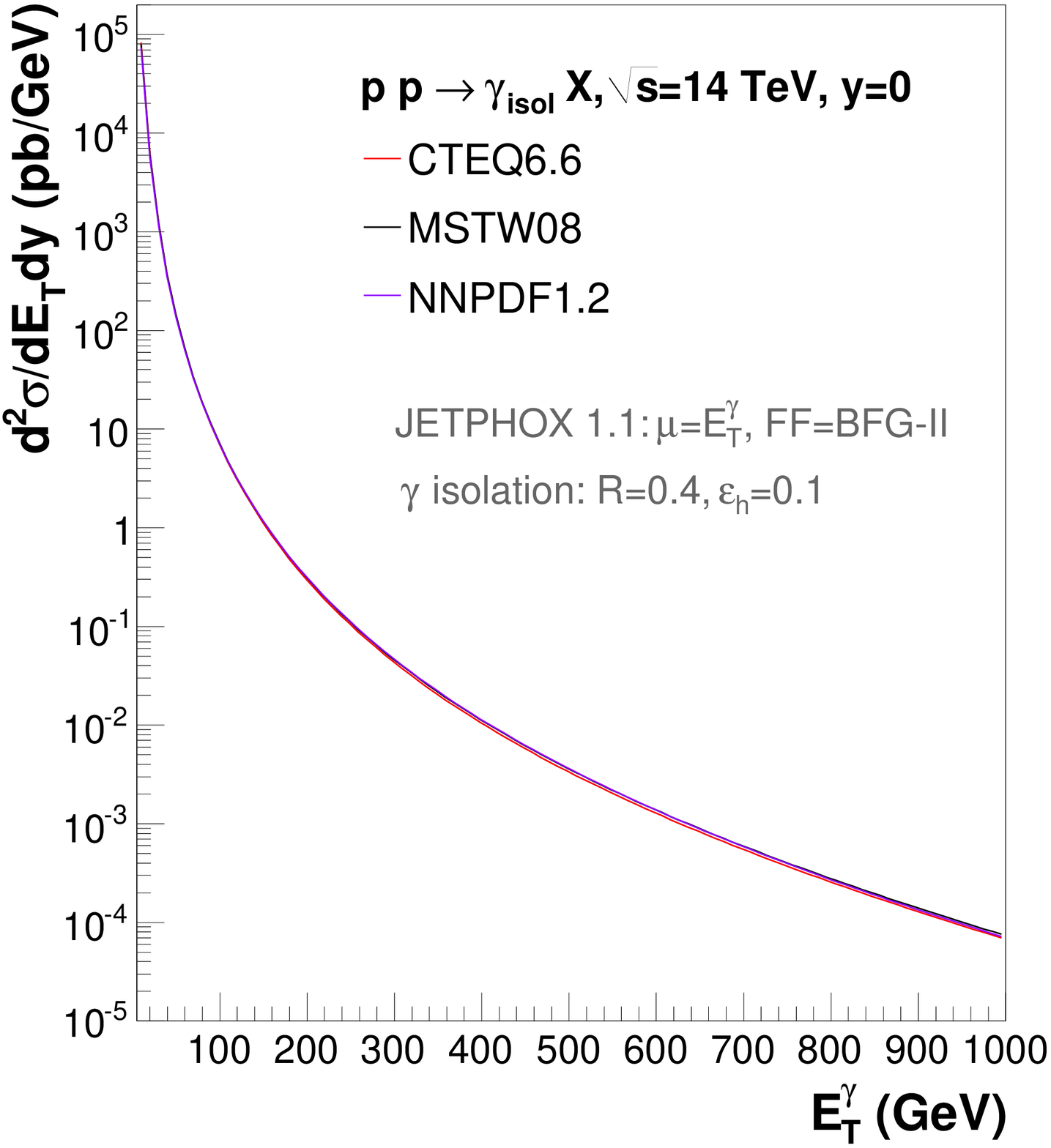}
\includegraphics[width=7.35cm,height=9.2cm]{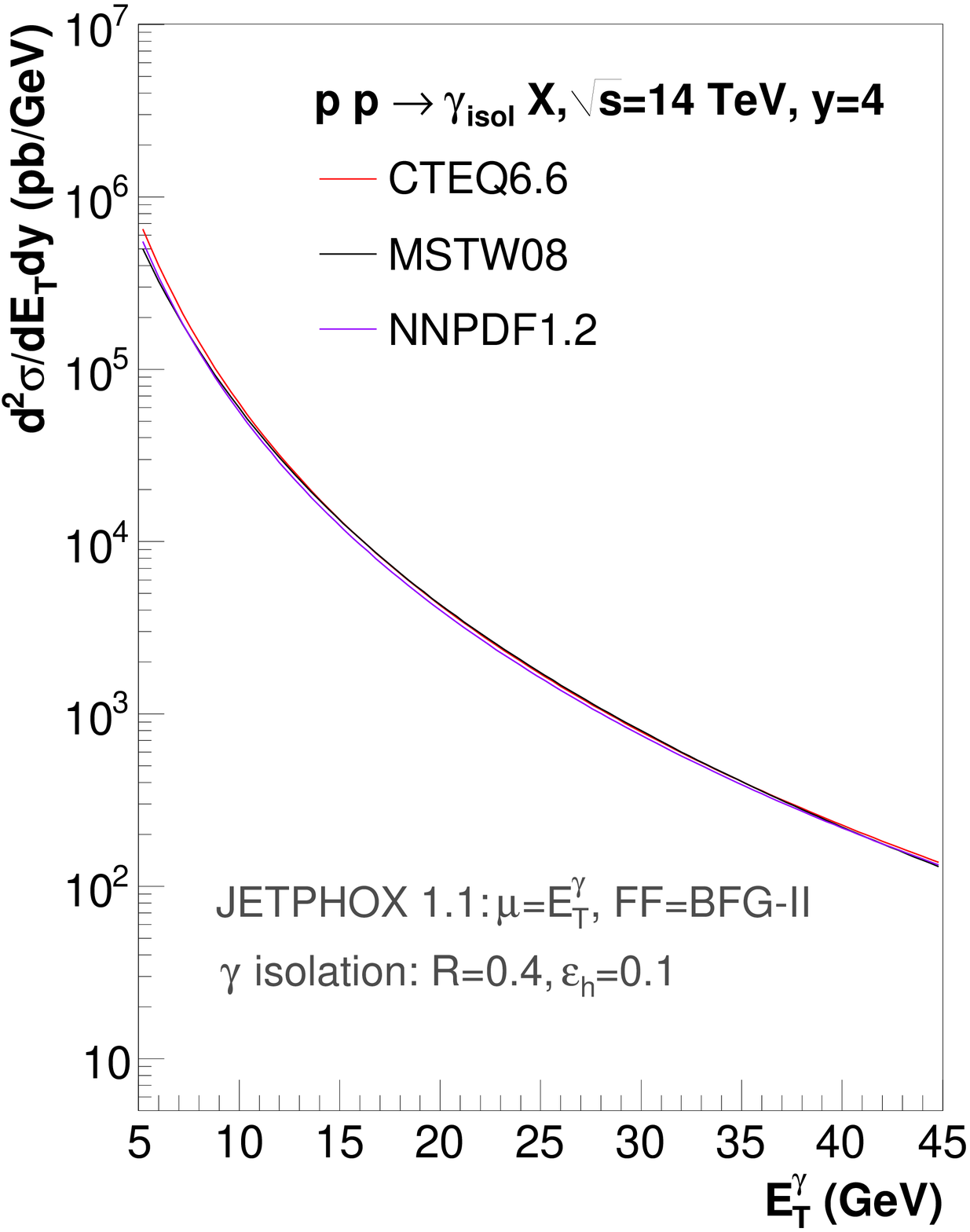}
\caption{Predictions of the isolated photon spectrum in \pp\ collisions at $\sqrts$~=~14~TeV
obtained with \jetphox\ (scales fixed at $\mu=\ETg$ and FFs set to BFG-II) for three
different PDF parametrisations (CTEQ6.6, MSTW08 and NNPDF1.2) for central ($y$~=~0, left) 
and forward ($y$~=~4, right) rapidities.}
\label{fig:lhc_isol_spec}
\end{figure}

At the top LHC center-of-mass energy of 14 TeV, the production cross section 
of isolated photons, $\sigma(\ETg>10~$GeV$)|_{y=0}$~=~300~nb at NLO, 
is seven times higher than at Tevatron, $\sigma(\ETg>10~$GeV$)|_{y=0}$~=~40~nb. This will allow one
to carry out high-statistics measurements even with moderate integrated luminosities, without (or with
minimal) pileup.
The ALICE, ATLAS, CMS and LHCb experiments at the LHC have photon reconstruction capabilities with 
electromagnetic calorimetry in various rapidity ranges. The two general-purpose high-luminosity detectors
ATLAS~\cite{atlas} and CMS~\cite{cms} cover\footnote{We do not consider here the possibilities covered 
by the ATLAS and CMS forward hadron calorimeters up to $|y|<$~5.} $|y|<$~3 from $E_{_T}\approx$~10~GeV up to very 
large $\ETg\approx$~1~TeV, whereas ALICE~\cite{alice} covers $|y|<$~0.7 in a range of moderate 
$\ETg\approx$~5~--~100~GeV. LHCb~\cite{lhcb}, on the other hand, can measure photons within 
$2 < \eta < 5$ albeit up to relatively moderate $E_{_T}\approx$~20~GeV, 
as the calorimeter dynamic range is mostly optimised for low energy photons coming from $B$-mesons 
radiative decays. As possible benchmark measurements, we will consider isolated (radius $R$~=~0.4) 
photon production at $y$~=~0 (for ALICE, ATLAS and CMS) and at $y$~=~4 (for LHCb\footnote{The 
chosen rapidity $y=4$, lower than the maximum LHCb $y=5$ value, accounts for the effect of the 
$R=0.4$ isolation radius and possible additional fiducial cuts.}). The \jetphox\ predictions for 
central and forward rapidities for three different PDF parametrizations (CTEQ6.6, MSTW08 and NNPDF1.2) 
are shown in Fig.~\ref{fig:lhc_isol_spec}. 

%\subsection{Isolated photons at mid-rapidity}
%\subsection{PDF sensitivity}

\begin{figure}[htbp]
\centering
\includegraphics[width=8.cm,height=7.cm]{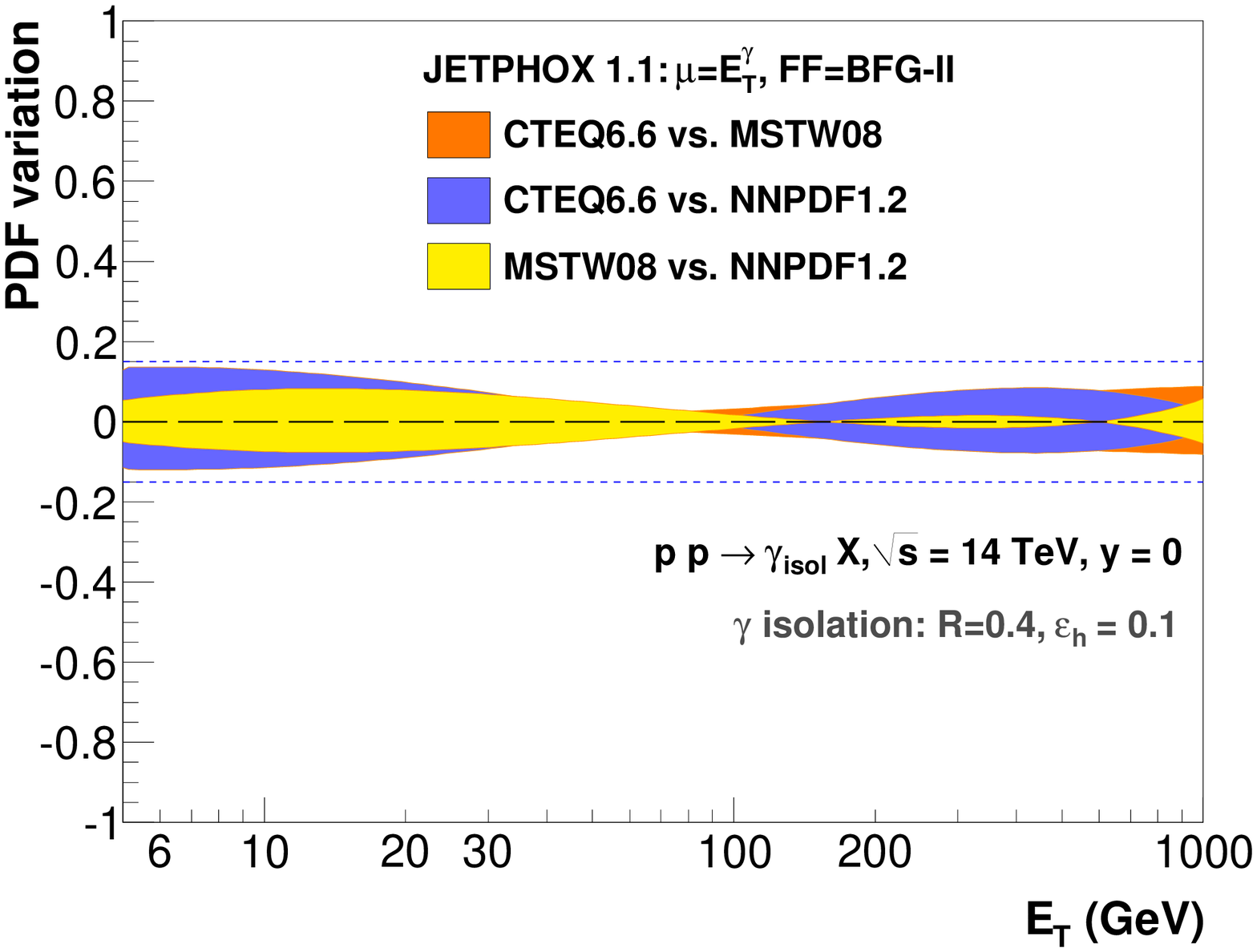}
\includegraphics[width=8.cm,height=7.cm]{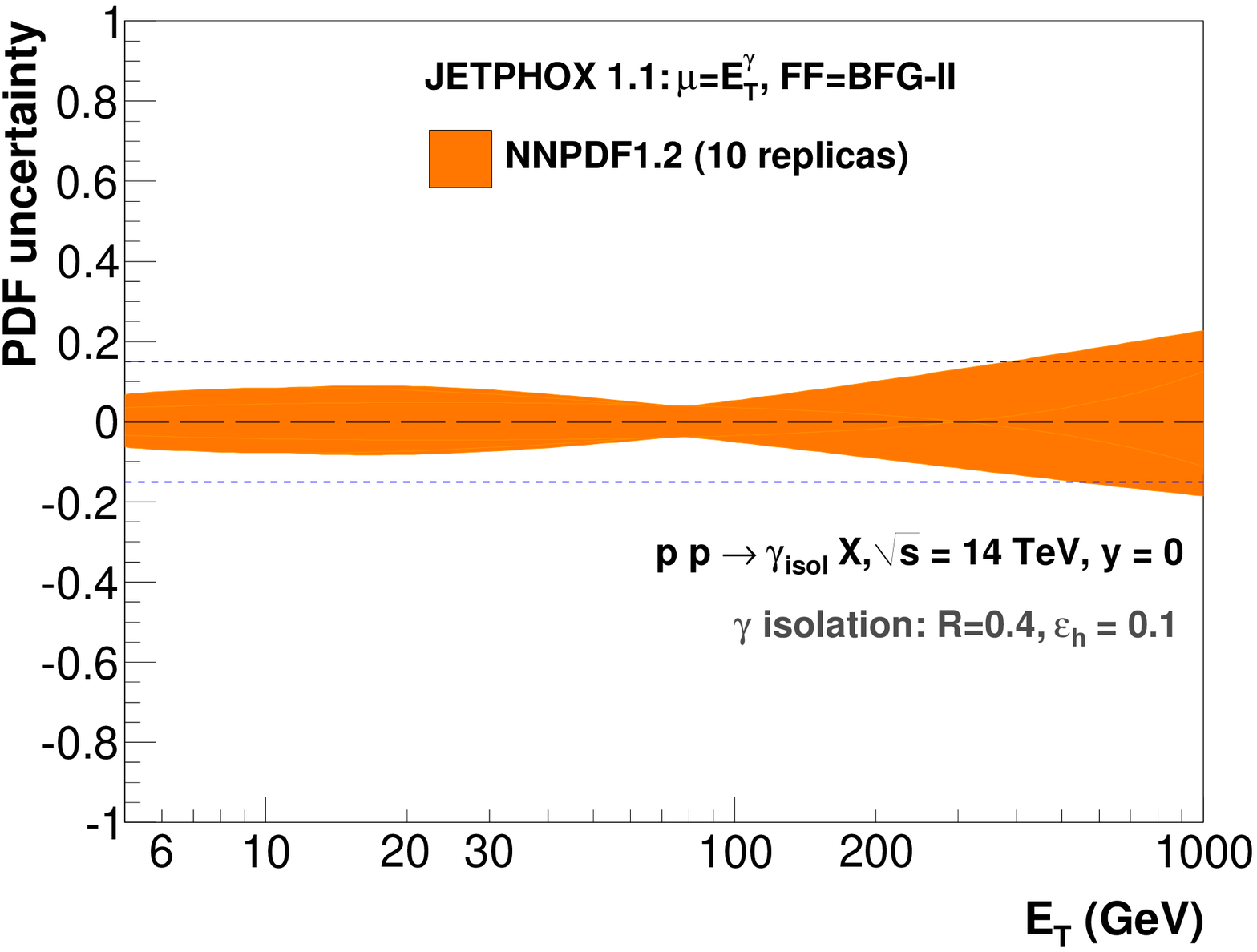}
\caption{Fractional differences between the isolated photon spectrum in \pp\ collisions at $\sqrts$~=~14~TeV 
at {\it central} rapidities ($y$~=~0) obtained with \jetphox\ (scales fixed at $\mu=\ETg$, FF set to BFG-II) using: 
(a) three different PDF parametrizations (CTEQ6.6, MSTW08 and NNPDF1.2), and (b) ten 
NNPDF1.2 replicas over the reference parametrization of the same PDF set. The dashed lines
indicate a $\pm$15\% uncertainty.}
\label{fig:lhc_isol_pdfs}
\end{figure}

Differences of a few tens of percent in the photon spectra obtained with different 
PDFs are not noticeable in logarithmic plots such as those in Fig.~\ref{fig:lhc_isol_spec}.
In Figure~\ref{fig:lhc_isol_pdfs} we show the fractional differences, $(1\pm$PDF1/PDF2), between the \jetphox\ predictions 
for isolated photon spectra ($R$~=~0.4 and $\varepsilon_h$~=~0.1) obtained at $y$~=~0 in \pp\ collisions 
at $\sqrts$~=~14~TeV for the various PDF sets. They have been obtained with the CTEQ66, MSTW08 and NNPDF1.2 
for fixed scales at $\mu = \ETg$ and BFG-II FF. 
%The left plot shows the relative differences between PDF sets. 
%The right plot gauges the PDF uncertainties within a {\it single} PDF set obtained comparing
%10 replicas of the NNPDF1.2 to the default NNPDF1.2 set. 
In general, the agreement among different PDFs is very good except at very low and very high $\ETg$. 
The choice of PDF results in up to $\sim$15\% variations of the photon yields below
$\ETg\approx$~15~GeV (left plot). Similar PDF uncertainties are obtained within one single 
PDF parametrization. As an example in the right plot we show the envelope of maximum relative differences 
at each $\ETg$ between each one of the ten\footnote{Although this is a relatively small number of
replicas, running \jetphox\ at NLO is a quite time-expensive operation from the computational point of view.} 
first NNPDF1.2 replicas and the default (central) NNPDF1.2 set.\\

%\begin{figure}[htbp]
%%\vspace{9pt}
%\centering
%\includegraphics[width=8.cm,height=8.5cm]{lhc_pp_isol_photons_jetphox_pdfall_mu1.eps}
%\includegraphics[width=8.cm,height=8.5cm]{lhc_pp_isol_photons_jetphox_pdfnnpdf_rep_mu1.eps}
%\caption{Predictions of the isolated photon spectrum in \pp\ collisions at $\sqrts$~=~14~TeV at $y$~=~0  
%obtained with \jetphox\ (scales fixed at $\mu=\ETg$, FF set to BFG-II) using (a) three different PDFs: 
%CTEQ6.6, MSTW08 and NNPDF1.2, and (b) 10 replicas of the NNPDF1.2 set (right).}
%\label{fig:lhc_isol_pdfs}
%\end{figure}

\begin{figure}[htbp]
\centering
\includegraphics[width=8.cm,height=7.cm]{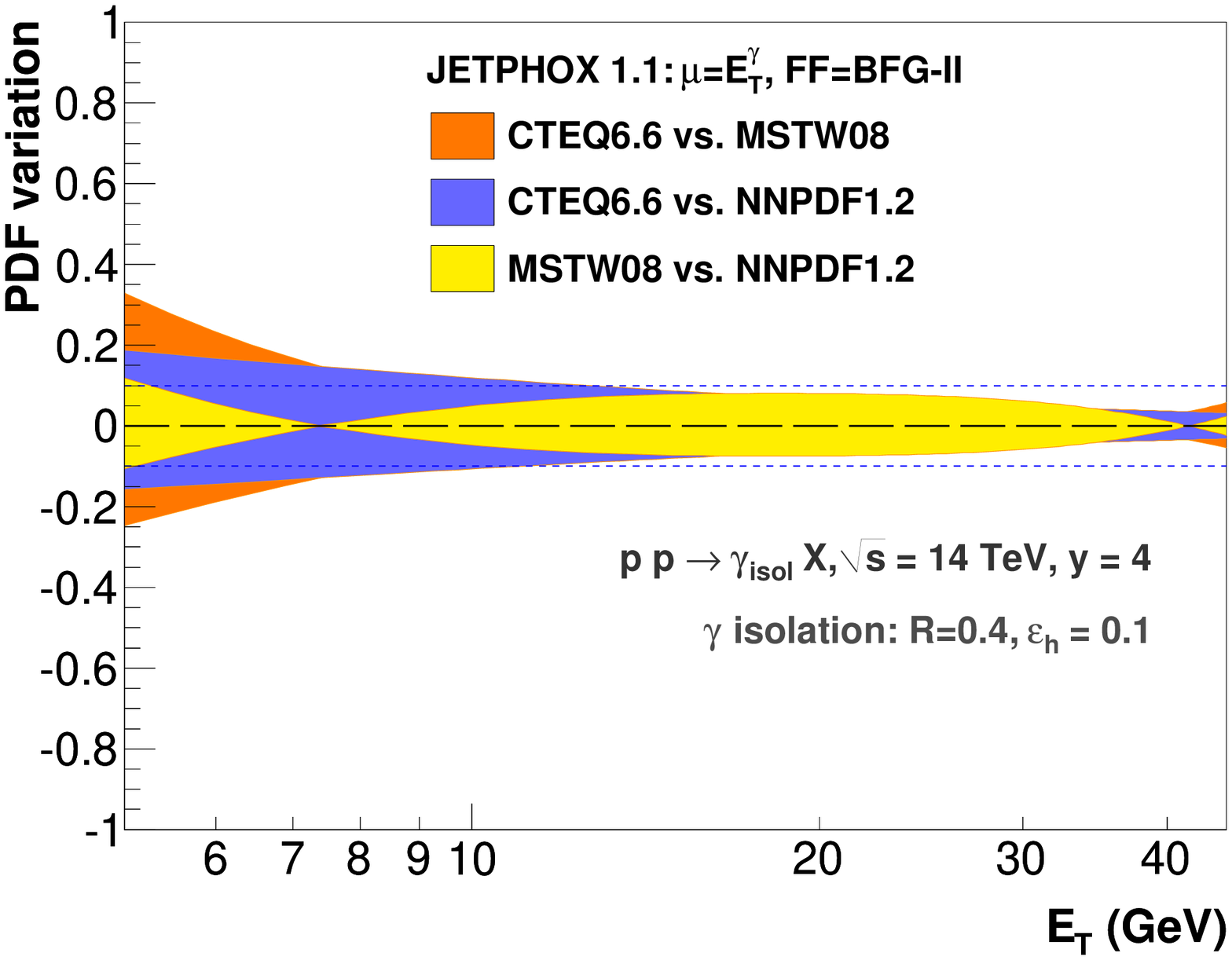}
\includegraphics[width=8.cm,height=7.cm]{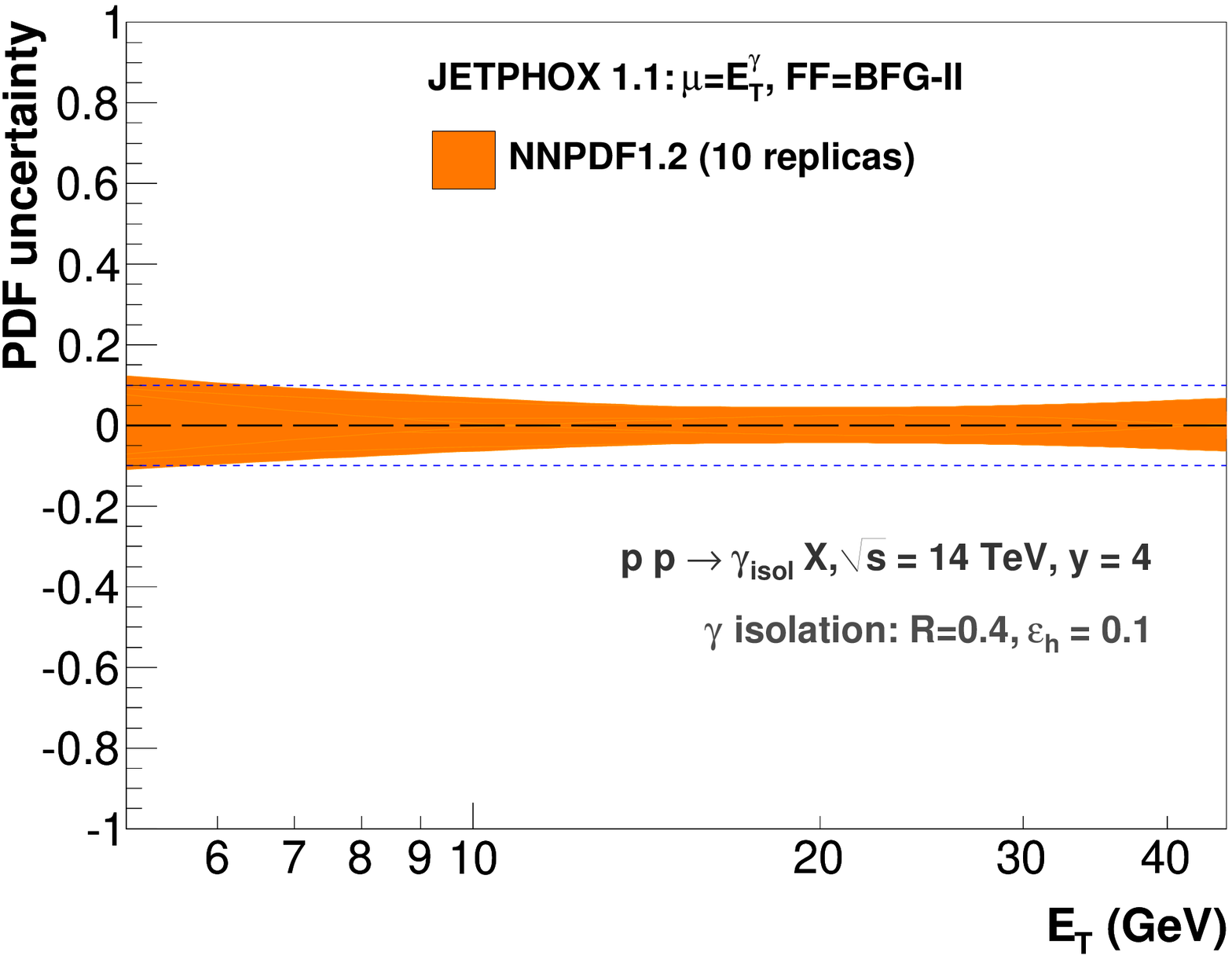}
\caption{Fractional differences between the isolated photon spectrum in \pp\ collisions at $\sqrts$~=~14~TeV 
at {\it forward} rapidities ($y$~=~4) obtained with \jetphox\ (scales fixed at $\mu=\ETg$, FF set to BFG-II) using: 
(a) three different PDF parametrizations (CTEQ6.6, MSTW08 and NNPDF1.2), and (b) ten 
NNPDF1.2 replicas over the reference parametrization of the same PDF set. The dashed 
lines indicate a $\pm$10\% uncertainty.}
\label{fig:lhcb_isol_pdfs}
\end{figure}

The expected PDF sensitivity for measurements of {\it forward} isolated photons ($y$~=~4) in the
transverse energy range $E_{_T} = 5 - 50$~GeV is shown in Figure~\ref{fig:lhcb_isol_pdfs}. 
Variations in the photon yields of up to $\sim$30\% depending on the PDF set can be seen at 
the lowest $\ETg\approx$~5~GeV (left plot). Intra-PDF uncertainties seem to be somewhat lower 
(around $\sim$15\% within 10 replicas of the NNPDF1.2 parametrization, right plot).

%\subsection{Isolated photons at forward rapidities}
%\subsection{Fragmentation function dependence}

%Figure~\ref{fig:lhcb_isol_pdfs} shows the prediction of \jetphox\ using CTEQ6.6, MSTW08 and NNPDF1.2 parametrizations.
%Figure~\ref{fig:ratio_lhcb_isol_pdfs} shows the prediction of \jetphox\ using 10 replicas of the NNPDF1.2 parametrization.

%\begin{figure}[htbp]
%%\vspace{9pt}
%\centering
%\includegraphics[width=8.cm,height=8.5cm]{lhcb_pp_isol_photons_jetphox_pdfall_mu1.eps}
%\includegraphics[width=8.cm,height=8.5cm]{lhcb_pp_isol_photons_jetphox_pdfnnpdf_rep_mu1.eps}
%\caption{Predictions of the isolated photon spectrum in \pp\ collisions at $\sqrts$~=~14~TeV at $y$~=~4
%obtained with \jetphox\ (scales fixed at $\mu=\ETg$, FF set to BFG-II) using (a) three different PDFs: CTEQ6.6, 
%MSTW08 and NNPDF1.2, and (b) 10 replicas of the NNPDF1.2 set (right).}
%\label{fig:lhcb_isol_pdfs}
%\end{figure}

%\subsection{Other uncertainties: theoretical scales and parton-to-photon FF}
%\subsection{Experimental and theoretical uncertainties}

As in the Tevatron case, the dominant source of experimental uncertainty (orange bands in 
Fig.~\ref{fig:pdf_dep}) in any (isolated or not) photon measurement at the LHC will likely come from the 
uncertainty in the absolute energy scale of the calorimeters. A typical 1.5\% uncertainty in the
photon energy calibration will propagate into about 10\% uncertainties in the photon cross
sections for a steeply-falling photon spectrum with inverse power-law exponent $n\approx$~5~--~8. 
Altogether, ten to fifteen percent uncertainties in the final cross sections (not accounting for 
an overall luminosity normalization error) are to be expected at the LHC. Those values are usually
better than the uncertainties linked to jet measurements where the energy-scale uncertainty, 
involving a good knowledge of the hadronic calorimeters calibration, is often poorer than for 
photons.\\

\begin{figure}[htbp]
%\vspace{9pt}
\centering
\includegraphics[width=8.cm,height=6.5cm]{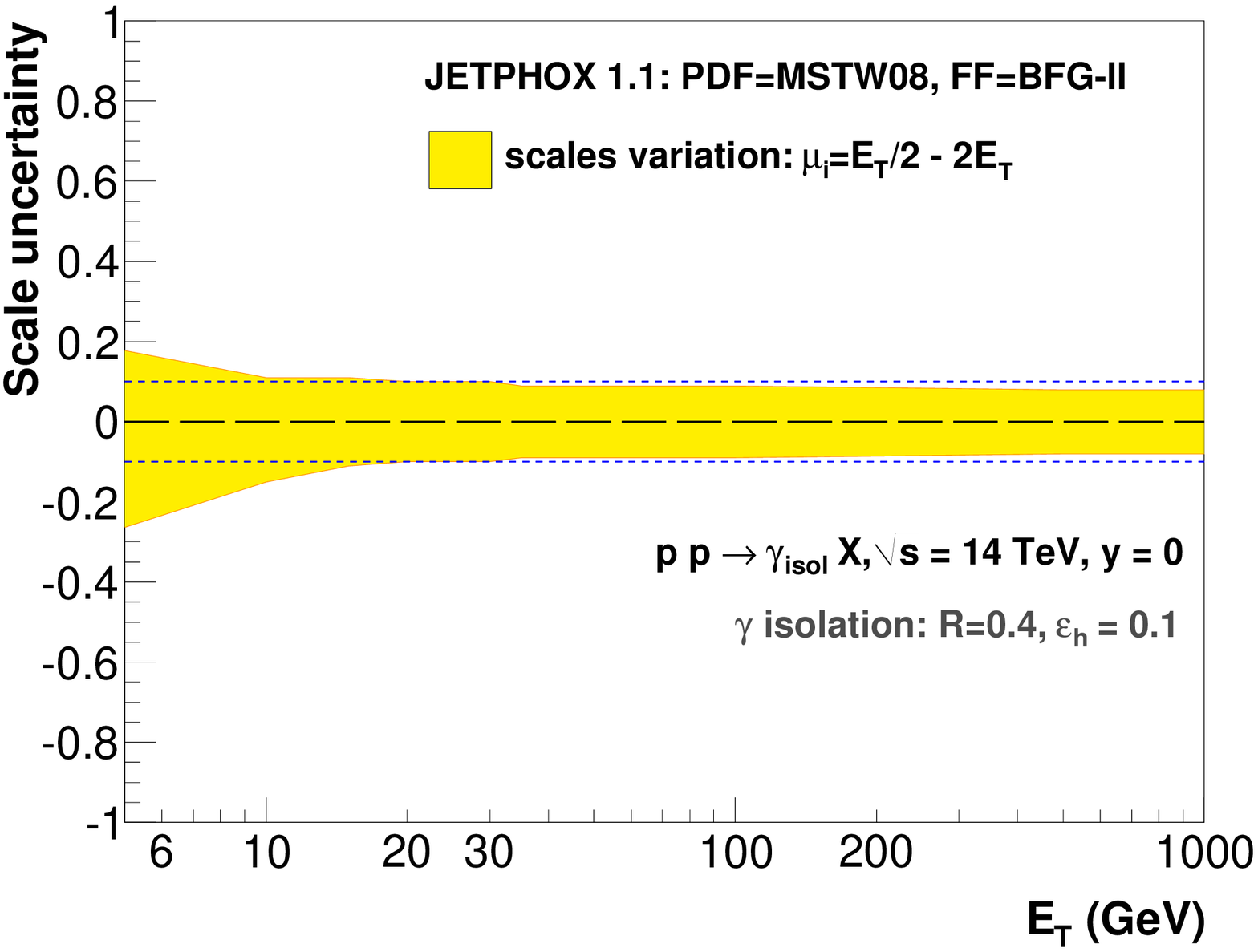}
\includegraphics[width=8.cm,height=6.5cm]{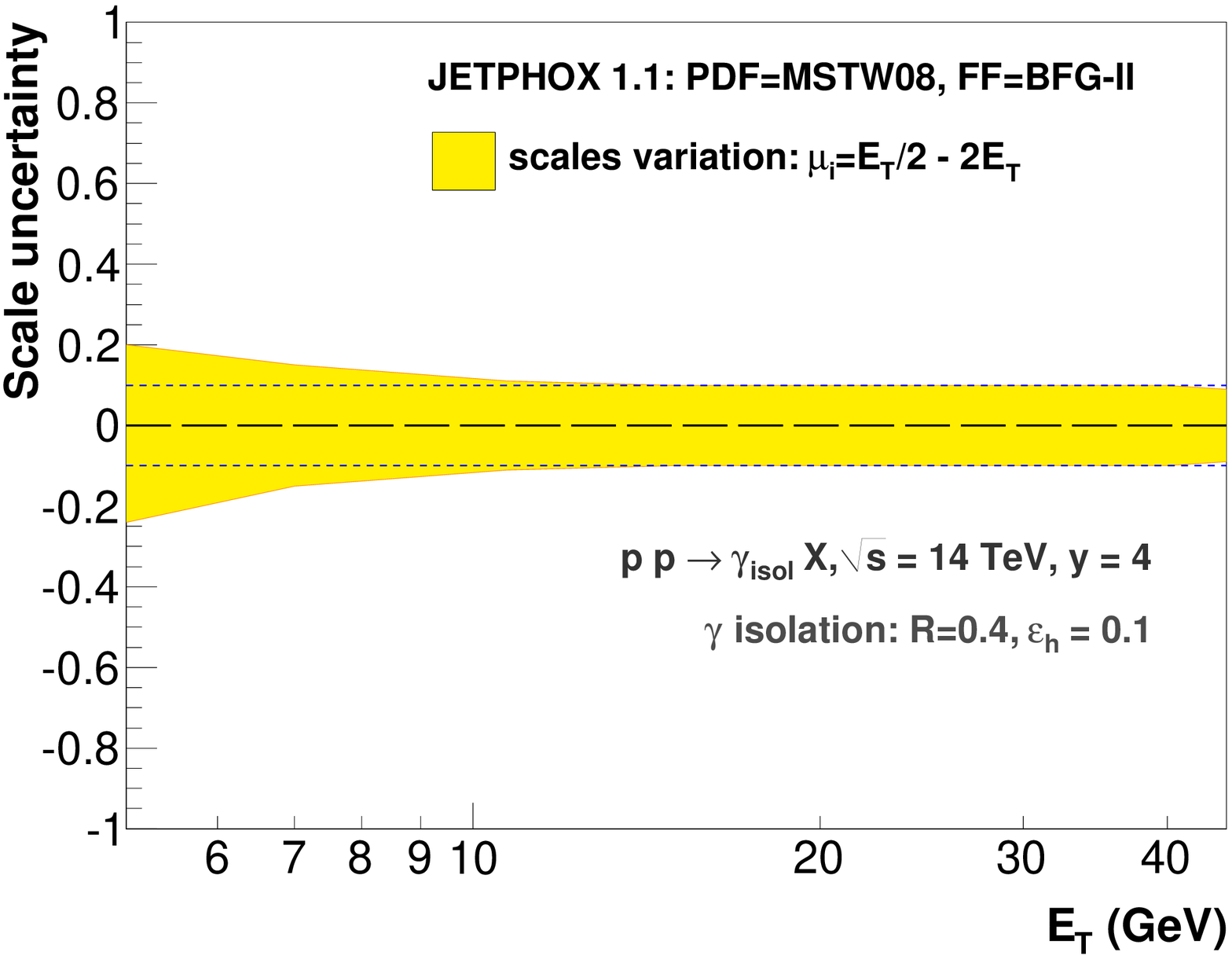}
\caption{Scale uncertainty plotted as the fractional differences between the
isolated photon $E_{_T}$-differential cross section in \pp\ at $\sqrts$~=~14~TeV 
at central ($y$~=~0, left plot) and forward ($y$~=~4, right plot) rapidities 
with scales $\mu=\ETg$ compared to that with scales: $\mu = 0.5\ETg, 2\ETg$. 
The dashed lines indicate a $\pm$10\% uncertainty.}
\label{fig:lhc_scale_dep}
\end{figure}

In the theoretical side, as seen in the case of the Tevatron predictions (Section~\ref{sec:data_th}), 
the choice of the parton-to-photon FF plays no role on the final 
isolated photon cross sections and we are left with the choice of theoretical scales
%In the low-$\ETg$ region where differences between the cross sections predicted with different 
%PDFs are the largest, the most important source of uncertainty is linked to the 
$\mu_{_R}$, $\mu_{_F}$ and $\mu_{_{ff}}$ as the only important source of uncertainty. 
In Figure~\ref{fig:lhc_scale_dep}, we present the fractional differences, $(1\pm \mu'/\mu)$, 
between the predicted isolated photon spectra obtained with \jetphox\ (MSTW08 PDFs and BFG-II FFs) with
``default'' theoretical scale ($\mu~=~\ETg$) and those obtained with $\mu'$~=~$\ETg/2$~--~$2\ETg$. 
At low $E_{_T}$ the uncertainty in the cross sections linked to scales variations amounts to up to $\pm$20\% 
%(resp. $\pm$30\%) at central (resp. forward) rapidities 
but it saturates at around $\pm$10\% above $\ETg\approx$~15~GeV at central and forward rapidities.\\
%The possible use of isolated photons measured at the LHC to constrain the proton parton densities will thus require
%a reduction of the scales dependence of the perturbative calculations. One venue to explore would be the
%inclusion of higher-order terms, e.g. in computations at NNLL accuracy~\cite{nll}, which reduces effectively 
%the scale dependence by a factor of two.

%%%%%%%%%%%%%%%%%%%%%%%%%%%%%%%%%%%%%%%%%%%%
%\clearpage

%\section{Discussion}
%\label{sec:disc}

As aforementioned, global PDF fits do not reliably determine the gluon for $x\lesssim 10^{-2}$ at low, 
yet perturbative, $Q^2\lesssim$~20~GeV$^2$ scales~\cite{Martin:2007sb}. This is due partly to the lack of 
precise structure function data for $x<10^{-4}$ and mainly due to the fact that the copious DIS $F_2$ 
results probe the quark distribution, while the gluon density is constrained by the evolution, $\log(Q^2)$ 
dependence, of these data. In the low $x$ region the available $Q^2$ range decreases since $x$ and $Q^2$ 
are correlated ($x \propto Q^2/s$) and the accuracy of the gluon determination becomes worse. 
The main motivation to use the LHC isolated photon spectra to help constrain the low-$x$ gluon density 
in the proton is summarized in the left plots of Figs.~\ref{fig:lhc_isol_pdfs} and \ref{fig:lhcb_isol_pdfs}. 
Different PDFs predict isolated photon yields which, at the lowest $\ETg$ values, differ by up to $\pm$30\%. 
Most of such differences arise from the different small-$x$ gluon densities implemented in the three 
PDF sets. Indeed, as seen in Fig.~\ref{fig:subproc_isol}, the Compton quark-gluon scattering clearly 
dominates isolated $\gamma$ production at the LHC and, as indicated by Fig.~\ref{fig:kinem}, the LHC 
data at forward (central) rapidities probes parton fractional momenta around $x=10^{-4}$ (resp. $x=10^{-2}$). 
It is not unexpected that the largest PDF-dependent variations of the yields are observed in the low $\ETg$ 
range since it is indeed at moderate virtualities that the gluon density is more uncertain. 
%: $g(x,Q^2)$ is particularly badly constrained for scales below $Q^2\approx$~20~GeV$^2$~\cite{Caola:2009iy}. 
The minimum photon $E_{_T}\approx$~5~GeV considered here is pretty optimistic from an experimental point-of-view 
-- the reconstruction and isolation of photons becomes more challenging with decreasing transverse energies -- 
but it still corresponds to relatively large scales of order $Q^2\approx$~25~GeV$^2$. Having the
possibility to measure isolated photons with lower, but still perturbative, momenta around $\ETg\approx$~3~GeV 
would provide even more stringent constraints on the low-$x$ gluon distributions.\\

One can take full advantage of the measured isolated photon results to better determine $g(x,Q^2)$, 
by including them in future global-fit analyses. Indeed, as seen in Fig.~\ref{fig:lhc_scale_dep}, 
the scale uncertainties at low $\ETg$ are relatively large ($\pm$20\%) and thus a simple direct spectral 
comparison of the experimental $\ETg$-differential cross sections to the theoretical predictions will not 
unambiguously indicate the preferred PDF parametrization. 
The latest works of the NNPDF collaboration~\cite{nnpdf2} use, for example, about 3500 experimental 
points to carry out their PDF fits, including DIS ($F_2$, $F_L$ and $F_2^{\rm charm}$) and 
proton-(anti)proton (Drell-Yan, vector-boson and jets) results. Yet, in the small-$x$ region the 
available experimental statistics is quite limited: just about 19 (126) data-points below $x=10^{-4}$ 
(resp. $x=10^{-3}$), among which only $F_L$ is directly sensitive to the gluon. In these regions 
where there are little or no experimental constraints on the gluon, the most important source of 
the PDF uncertainty is due to the parametrization bias. Inclusion of new data-sets is basic to 
improve the central value of the PDFs
%cross-check the consistency of previous data, and reduce the associated PDF uncertainties. 
and reduce their associated uncertainties. 
In that respect, including 10~--~20 data-points from forward isolated photon spectra at $\sqrts$~=~14~TeV, 
will place new direct experimental constraints on the gluon distribution at $x=10^{-4}$.
The effect of these data on future PDF parametrizations might be larger if the existing measurements 
have inconsistencies which previously have not been accounted for.
%The somewhat disappointing result of our studies is that, if the differences between the central
%values of the various PDF parametrizations constitute a good proxy for the actual uncertainty on
%the gluon distribution, the LHC isolated photon data seem to provide only a moderate constrain on 
%$g$. Indeed, the three PDFs studied yield central values of 
%In the regions where there are little or no experimental constraints, uncertainty larger due to minimum
%parametrization bias. Reduction of uncertainties with respect to older NNPDF sets due to inclusion of new data.
%Uncertainties on PDFs competitive with results from other groups. Smaller uncertainty due to wide set of
%consistent data
%Adding photon data would thus provide a very stringent cross-check of QCD factorization, and its effect 
%might be larger if other measurements have inconsistencies which previously have not been accounted for
All in all, this work indicates that the combined photon data at all energies -- about 350 data-points 
collected so far plus $\cO{100}$ new data-points expected from the four experiments at the LHC -- 
provide new useful constraints of the gluon distribution in a wide ($x,Q^2$) range.

%%%%%%%%%%%%%%%%%%%%%%%%%%%%%%%%%%%%%%%%%%%%
%\clearpage

\section{Conclusions}% and outlook}
\label{sec:concl}

We have compared the most recent measurements of isolated photon spectra in \ppbar\ collisions
at $\sqrts$~=~1.96~TeV~\cite{d0,cdf} with next-to-leading-order calculations with updated parton 
distributions functions (PDFs) in the proton. Previous data-theory comparisons~\cite{Aurenche:2006vj,d0,cdf} 
used relatively older PDF sets. The $E_{_T}$-differential spectra at Tevatron can be equally well 
reproduced by the CTEQ6.6, MSTW08 and NNPDF1.2 global fit parametrizations within the existing 
experimental and theoretical uncertainties. Differences in the isolated photon cross sections 
for different PDF sets are in the range 5\%~--~10\%, with the MSTW08 spectrum being somewhat in between 
the spectra computed with the two other PDFs. Theoretical uncertainties linked to the scales choices
$\mu_i=\ETg/2 - 2\ETg$ result in around $\pm$10\% $\ETg$-independent changes in the NLO cross sections. 
The choice of parton-to-photon fragmentation function (FF) does not introduce any additional 
theoretical uncertainty as the isolation criteria help to keep the contribution of fragmentation photons 
to the total prompt photon yield well below the fifteen percent at all transverse energies, and differences 
between FFs are a few percent of this remaining 15\% contribution.\\

We have also presented the NLO isolated photon spectra expected in \pp\ collisions at the top
LHC energy for central ($y~=~0$) and forward ($y~=~4$) rapidities and determined the associated theoretical 
uncertainties. At $\sqrts$~=~14~TeV, the prompt photon cross sections (above 10 GeV) are more than 
a factor of seven larger than at Tevatron, and the parton fractional momenta probed are relatively small, 
down to $x\approx$~10$^{-3}$ (resp. 10$^{-5}$) at mid (resp. forward) rapidity. Since the quark-gluon 
Compton process is found to represent about two thirds of the total isolated photon cross section at 
the LHC for standard isolation cuts, such a measurement promises to provide an interesting direct 
measurement to the relatively unconstrained low-$x$ gluon density $g(x,Q^2)$ in the proton. The three 
PDFs studied yield central values of the photon spectrum which differ at most by $\pm$15\% at $y = 0$ 
and by $\pm$30\% at $y=4$ in the low-$\ETg$ region of the spectra, whereas the experimental
% Experimentally, the dominant uncertainties $\cO{10\%}$ will be 
(associated with the calibration of the energy scale in the electromagnetic calorimeters)
and the theoretical (linked to the choice of the  renormalisation, factorisation, and fragmentation scales)
uncertainties are of $\cO{10-20\%}$. %also of $\pm 10\%$.
To finalize, we have presented arguments to motivate the inclusion of the LHC photon data in future 
PDF global-fit analyses.\\

In summary, given (i) the high-quality isolated photon measurements available or expected at collider 
energies, (ii) the good agreement between the existing RHIC and Tevatron (except for the lowest $\ETg$ bins) 
data and NLO calculations, and (iii) the large statistics photon data expected to be collected at various 
rapidities at the LHC, it is worth to reconsider the inclusion of the combined photon collider data measured 
in proton-(anti)proton collisions -- about 350 data-points collected so far plus 
$\cO{100}$ new data-points expected from the four experiments at the LHC -- in global-fit PDF analyses. 
Such data provide direct extra constraints on the gluon PDF in a wide ($x,Q^2$) range.

%%%%%%%%%%%%%%%%%%%%%%%%%%%%%%%%%%%%%%%%%%%%
\clearpage

\section*{Acknowledgments}

We are grateful to Fran\c{c}ois Arleo for many discussions and a careful reading of previous
versions of this manuscript, as well as to Joan Rojo for valuable feedback on the NNPDF parton
distributions and for useful comments on the document. Discussions with Thierry Gousset
are also acknowledged. DdE is partially supported by the 7th EU Framework Programme 
(contract FP7-ERG-2008-235071).

%%%%%%%%%%%%%%%%%%%%%%%%%%%%%%%%%%%%%%%%%%%%
%\clearpage

%%%%%%%%%%%%%%%%%%%%%%%%%%%%%%%%%%%%%%%%%%%%

\end{document}